\PassOptionsToPackage{unicode}{hyperref}
\PassOptionsToPackage{hyphens}{url}
\documentclass[
  submission,
  copyright,
  creativecommons]{eptcs}
\usepackage{xcolor}
\usepackage{amsmath,amssymb}
\usepackage{iftex}
\ifPDFTeX
  \usepackage[T1]{fontenc}
  \usepackage[utf8]{inputenc}
  \usepackage{textcomp} 
\else 
  \usepackage{unicode-math} 
  \defaultfontfeatures{Scale=MatchLowercase}
  \defaultfontfeatures[\rmfamily]{Ligatures=TeX,Scale=1}
\fi
\usepackage{lmodern}
\ifPDFTeX\else
\fi
\IfFileExists{upquote.sty}{\usepackage{upquote}}{}
\IfFileExists{microtype.sty}{
  \usepackage[]{microtype}
  \UseMicrotypeSet[protrusion]{basicmath} 
}{}
\makeatletter
\@ifundefined{KOMAClassName}{
  \IfFileExists{parskip.sty}{%
    \usepackage{parskip}
  }{
    \setlength{\parindent}{0pt}
    \setlength{\parskip}{6pt plus 2pt minus 1pt}}
}{
  \KOMAoptions{parskip=half}}
\makeatother
\usepackage{color}
\usepackage{fancyvrb}

\DefineVerbatimEnvironment{Highlighting}{Verbatim}{commandchars=\\\{\}}
\newenvironment{Shaded}{}{}

\newcommand{\CommentTok}[1]{\textcolor[rgb]{0.38,0.63,0.69}{\textit{#1}}}

\newcommand{\DataTypeTok}[1]{\textcolor[rgb]{0.56,0.13,0.00}{#1}}
\newcommand{\DecValTok}[1]{\textcolor[rgb]{0.25,0.63,0.44}{#1}}

\newcommand{\KeywordTok}[1]{\textcolor[rgb]{0.00,0.44,0.13}{\textbf{#1}}}
\newcommand{\NormalTok}[1]{#1}

\newcommand{\OtherTok}[1]{\textcolor[rgb]{0.00,0.44,0.13}{#1}}
\newcommand{\PreprocessorTok}[1]{\textcolor[rgb]{0.74,0.48,0.00}{#1}}

\NewDocumentCommand\citeproctext{}{}

\makeatletter
 \let\@cite@ofmt\@firstofone
 \def\@biblabel#1{}
 \def\@cite#1#2{{#1\if@tempswa , #2\fi}}
\makeatother
\newlength{\cslhangindent}
\newlength{\csllabelwidth}
\newenvironment{CSLReferences}[2] 
 {\begin{list}{}{%
  \setlength{\itemindent}{0pt}
  \setlength{\leftmargin}{0pt}
  \setlength{\parsep}{0pt}
  \ifodd #1
   \setlength{\leftmargin}{\cslhangindent}
   \setlength{\itemindent}{-1\cslhangindent}
  \fi
  \setlength{\itemsep}{#2\baselineskip}}}
 {\end{list}}
\usepackage{calc}

\providecommand{\event}{ACT 2026}

\AtBeginDocument{\author{C.B. Aberl\'{e}
\institute{Carnegie Mellon University}
\email{caberle@andrew.cmu.edu}
}}

\newcommand{\titlerunning}{Compositional Program Verification with Polynomial Functors}
\newcommand{\authorrunning}{C.B. Aberl\'{e}}

\hypersetup{
  bookmarksnumbered,
  pdftitle  = {\titlerunning},
  pdfauthor = {\authorrunning},
  pdfsubject = {EPTCS},
  pdfkeywords = {polynomial functors, dependent types, compositional verification, Mealy machines, Agda}
}

\AddToHook{begindocument/end}{\setcounter{secnumdepth}{3}}

\usepackage{etoolbox}
\newcommand{\restorethmtopsep}{%
  \topsep=8pt plus 2pt minus 2pt\relax
  \partopsep=0pt\relax}
\AddToHook{begindocument/end}{%
  \BeforeBeginEnvironment{theorem}{\restorethmtopsep}%
  \BeforeBeginEnvironment{proposition}{\restorethmtopsep}%
  \BeforeBeginEnvironment{lemma}{\restorethmtopsep}%
  \BeforeBeginEnvironment{corollary}{\restorethmtopsep}%
  \BeforeBeginEnvironment{definition}{\restorethmtopsep}%
  \BeforeBeginEnvironment{example}{\restorethmtopsep}%
  \BeforeBeginEnvironment{remark}{\restorethmtopsep}%
}

\usepackage{amsthm}
\newtheorem{theorem}{Theorem}[section]
\newtheorem{proposition}[theorem]{Proposition}

\theoremstyle{definition}
\newtheorem{definition}[theorem]{Definition}

\theoremstyle{remark}

\usepackage{mathpartir}
\usepackage{quiver}

\usepackage[utf8]{inputenc}
\usepackage{newunicodechar}

\newunicodechar{λ}{\ensuremath{\lambda}}
\newunicodechar{Π}{\ensuremath{\Pi}}
\newunicodechar{Σ}{\ensuremath{\Sigma}}
\newunicodechar{∘}{\ensuremath{\circ}}
\newunicodechar{∀}{\ensuremath{\forall}}
\newunicodechar{→}{\ensuremath{\to}}
\newunicodechar{≡}{\ensuremath{\equiv}}
\newunicodechar{ℓ}{\ensuremath{\ell}}
\newunicodechar{₀}{\ensuremath{_0}}
\newunicodechar{₁}{\ensuremath{_1}}
\newunicodechar{₂}{\ensuremath{_2}}
\newunicodechar{₃}{\ensuremath{_3}}
\newunicodechar{₄}{\ensuremath{_4}}
\newunicodechar{₅}{\ensuremath{_5}}
\newunicodechar{⊔}{\ensuremath{\sqcup}}
\newunicodechar{⊤}{\ensuremath{\top}}
\newunicodechar{≐}{\ensuremath{\doteq}}
\newunicodechar{κ}{\ensuremath{\kappa}}
\newunicodechar{Δ}{\ensuremath{\Delta}}
\newunicodechar{Λ}{\ensuremath{\Lambda}}
\newunicodechar{δ}{\ensuremath{\delta}}
\newunicodechar{≤}{\ensuremath{\leq}}
\newunicodechar{∈}{\ensuremath{\in}}
\newunicodechar{∩}{\ensuremath{\cap}}
\newunicodechar{¬}{\ensuremath{\neg}}
\newunicodechar{∧}{\ensuremath{\wedge}}
\newunicodechar{∨}{\ensuremath{\vee}}
\newunicodechar{β}{\ensuremath{\beta}}
\newunicodechar{⊥}{\ensuremath{\bot}}
\newunicodechar{⊕}{\ensuremath{\oplus}}
\newunicodechar{⊙}{\ensuremath{\odot}}
\newunicodechar{⇒}{\ensuremath{\Rightarrow}}
\newunicodechar{×}{\ensuremath{\times}}
\newunicodechar{∣}{\ensuremath{\mid}}
\newunicodechar{⊗}{\ensuremath{\otimes}}
\usepackage{bookmark}
\IfFileExists{xurl.sty}{\usepackage{xurl}}{} 
\hypersetup{
  pdftitle={Compositional Program Verification with Polynomial Functors in Dependent Type Theory},
  hidelinks,
  pdfcreator={LaTeX via pandoc}}

\title{Compositional Program Verification with Polynomial Functors in
Dependent Type Theory}
\author{}
\date{}

\begin{document}
\maketitle
\begin{abstract}
We present a framework for compositional program verification based on
polynomial functors in dependent type theory. In this framework,
polynomial functors serve as program interfaces, Kleisli morphisms for
the free monad monad serve as implementations, and dependent polynomials
encode pre/postcondition specifications. We show that implementations
and their verifications compose via wiring diagrams, and that Mealy
machines provide a compositional coalgebraic operational semantics. We
identify the abstract categorical structure underlying this
compositionality as a monoidal functor from specifications to interfaces
with a compatible monoidal natural transformation of lax monoidal
presheaves; this opens the door to generalizations to other categories,
monoidal products, etc., including settings for concurrency and
relational verification, which we sketch. As a proof-of-concept, the
entire framework has been formalized in Agda.
\end{abstract}

\section{Introduction}\label{introduction}

Large software systems are often opaque; their capabilities need not be.
To verify the behavior of complex programs, we ought to be able to
decompose them into simpler components, verify those components
independently, and compose the results. In this paper, we develop such a
\emph{compositional} approach to program verification using
\emph{polynomial functors} in dependent type theory. A polynomial
functor \(P(y) = \sum_{a : A} y^{B(a)}\) naturally represents the
interface of a dependent function \((a : A) \to B(a)\), and a morphism
to the \emph{free monad} on another polynomial represents a program
module implementing one interface by calling another. These
constructions generalize to \emph{dependent polynomials} encoding
pre/postcondition specifications, allowing programs to be verified in
exactly the same manner they are built up.

This perspective connects to work on polynomial functors as a theory of
interaction (Niu and Spivak 2024; Libkind and Spivak 2025; Spivak 2022).
Our contribution is to extend this picture to dependent type theory,
introducing dependent polynomials that encode program specifications,
and showing that the resulting verification framework is fully
compositional. Specifically, we show that implementations and
verifications compose along \emph{wiring diagrams} (\S 2--3); that
\emph{Mealy machines} provide a compatible coalgebraic operational
semantics (\S 4); that \emph{dependent polynomials} and \emph{dependent
free monads} yield compositional verification (\S 5--6); and that the
abstract categorical structure underlying this compositionality can be
identified as a monoidal functor with compatible lax monoidal presheaves
(\S 7) connected by a monoidal natural transformation. We also sketch an
extension to concurrent modules via a parallel monoidal product
(Appendix A). The entire development has been formalized in Agda (Norell
2009).

\section{Polynomial Functors and Program
Interfaces}\label{polynomial-functors-and-program-interfaces}

\subsection{Polynomial Functors}\label{polynomial-functors}

A \emph{polynomial functor} is an endofunctor on the category of sets
(or, in our case, types) of the form \[P(y) = \sum_{a : A} y^{B(a)}\]
where \(A\) is a type and \(B : A \to \mathsf{Set}\) is a family of
types indexed by \(A\). We call \(A\) the type of \emph{positions} and
\(B(a)\) the type of \emph{directions} at position \(a\) (Niu and Spivak
2024). Concretely, we represent a polynomial functor as a pair
\((A, B)\):

\begin{Shaded}
\begin{Highlighting}[]
\NormalTok{Poly }\OtherTok{:} \DataTypeTok{Set₁}
\NormalTok{Poly }\OtherTok{=}\NormalTok{ Σ }\DataTypeTok{Set} \OtherTok{(λ}\NormalTok{ A }\OtherTok{→}\NormalTok{ A }\OtherTok{→} \DataTypeTok{Set}\OtherTok{)}
\end{Highlighting}
\end{Shaded}

Given a polynomial \(p = (A, B)\) and a functor
\(F : \mathsf{Set} \to \mathsf{Set}\), the type of \emph{morphisms}
(natural transformations) from \(p\) to \(F\) is:

\begin{Shaded}
\begin{Highlighting}[]
\OtherTok{\_}\NormalTok{⇒}\OtherTok{\_} \OtherTok{:}\NormalTok{ Poly }\OtherTok{→} \OtherTok{(}\DataTypeTok{Set} \OtherTok{→} \DataTypeTok{Set}\OtherTok{)} \OtherTok{→} \DataTypeTok{Set}
\OtherTok{(}\NormalTok{A , B}\OtherTok{)}\NormalTok{ ⇒ F }\OtherTok{=} \OtherTok{(}\NormalTok{x }\OtherTok{:}\NormalTok{ A}\OtherTok{)} \OtherTok{→}\NormalTok{ F }\OtherTok{(}\NormalTok{B x}\OtherTok{)}
\end{Highlighting}
\end{Shaded}

A morphism \(p \Rightarrow q\) between two polynomials, where
\(p = (A, B)\) and \(q = (C, D)\), thus consists of a \emph{forward} map
\(f_0 : A \to C\) on positions and a \emph{backward} map
\(f^1 : (a : A) \to D(f_0(a)) \to B(a)\) on directions. In the language
of the polynomial functors literature, this is a \emph{lens} from \(p\)
to \(q\) (Niu and Spivak 2024, Ch. 3).

Thinking of \((A, B)\) as the interface of a dependent function
\(g : (a : A) \to B(a)\), and \((C, D)\) as the interface of a function
\(h : (c : C) \to D(c)\), a lens from \((A,B)\) to \((C,D)\) encodes a
program that implements \(g\) by calling \(h\): given input \(a : A\),
it calls \(h\) on \(f_0(a)\), then applies \(f^1(a)\) to the result to
produce an output of type \(B(a)\). This is the basic pattern of
\emph{assume-guarantee reasoning} that we shall see throughout the
paper: assuming an implementation of \(h\), we guarantee an
implementation of \(g\).

\subsection{Free Monads and Sequential
Composition}\label{free-monads-and-sequential-composition}

The notion of morphism/lens described above only captures situations
where the assumed function is called \emph{exactly once}. To express
more general patterns of interaction, where the assumed interface may be
called zero, one, or many times---with each subsequent call potentially
depending on the results of previous ones---we use the \emph{free monad}
on a polynomial functor (Niu and Spivak 2024; Libkind and Spivak 2025):

\begin{Shaded}
\begin{Highlighting}[]
\KeywordTok{data}\NormalTok{ Free }\OtherTok{(}\NormalTok{p }\OtherTok{:}\NormalTok{ Poly}\OtherTok{)} \OtherTok{(}\NormalTok{C }\OtherTok{:} \DataTypeTok{Set}\OtherTok{)} \OtherTok{:} \DataTypeTok{Set} \KeywordTok{where}
\NormalTok{    return }\OtherTok{:}\NormalTok{ C }\OtherTok{→}\NormalTok{ Free p C}
\NormalTok{    bind }\OtherTok{:} \OtherTok{(}\NormalTok{x }\OtherTok{:}\NormalTok{ p }\OtherTok{.}\NormalTok{fst}\OtherTok{)} \OtherTok{→} \OtherTok{(}\NormalTok{p }\OtherTok{.}\NormalTok{snd x }\OtherTok{→}\NormalTok{ Free p C}\OtherTok{)} \OtherTok{→}\NormalTok{ Free p C}
\end{Highlighting}
\end{Shaded}

An element of \(\mathsf{Free}\,p\,C\) is a computation that either
returns a value of type \(C\), or makes a call to \(p = (A, B)\) with
some input \(x\), receives a response, and continues depending on the
result---equivalently, such programs correspond to well-founded trees
with leaves in \(C\) (Libkind and Spivak 2025; Niu and Spivak 2024) and
internal nodes labeled by positions \(a : A\), with branches indexed by
directions \(b : B(a)\).

As the name would imply, the free monad carries a monadic structure,
with \texttt{bind} sequencing computations. We use Agda's
\texttt{syntax} mechanism to introduce notations as follows:
\texttt{call{[}\ b\ ←\ a\ {]}\ e} makes a single call with input
\texttt{a}, binds the result to \texttt{b}, and continues with
\texttt{e}; \texttt{do{[}\ b\ ←\ m\ {]}\ e} sequences a sub-computation
\texttt{m} similarly (the reader familiar with monadic programming in
languages such as Haskell will recognize these patterns). These are used
throughout the paper.

This brings us to our central definition for program modules:

\begin{definition}\label{def:implementation}
An \emph{implementation} of interface $p$ depending on interface $q$ is a morphism $p \Rightarrow \mathsf{Free}\,q$, i.e., a function that, given any input to $p$, produces a computation in the free monad on $q$ that returns an appropriate output.
\end{definition}

\subsection{Sums of Polynomial
Functors}\label{sums-of-polynomial-functors}

In practice, program modules and their interfaces often involve
\emph{multiple} functions. We handle this via the \emph{sum} of
polynomial functors. Given an indexing type \(U\) and a family
\(P : U \to \mathsf{Poly}\), their sum \(\sum_{u:U} P(u)\) is again a
polynomial, which is in fact the coproduct in the category
\(\mathbf{Poly}\) of polynomial functors and their morphisms:

\begin{Shaded}
\begin{Highlighting}[]
\NormalTok{sum }\OtherTok{:} \OtherTok{(}\NormalTok{U }\OtherTok{:} \DataTypeTok{Set}\OtherTok{)} \OtherTok{→} \OtherTok{(}\NormalTok{p }\OtherTok{:}\NormalTok{ U }\OtherTok{→}\NormalTok{ Poly}\OtherTok{)} \OtherTok{→}\NormalTok{ Poly}
\NormalTok{sum U p }\OtherTok{.}\NormalTok{fst }\OtherTok{=}\NormalTok{ Σ U }\OtherTok{(λ}\NormalTok{ u }\OtherTok{→}\NormalTok{ p u }\OtherTok{.}\NormalTok{fst}\OtherTok{)}
\NormalTok{sum U p }\OtherTok{.}\NormalTok{snd }\OtherTok{(}\NormalTok{u , x}\OtherTok{)} \OtherTok{=}\NormalTok{ p u }\OtherTok{.}\NormalTok{snd x}
\end{Highlighting}
\end{Shaded}

A morphism \(\mathsf{sum}\,U\,P \Rightarrow \mathsf{Free}\,q\) then
amounts to a family of implementations, one for each label \(u : U\). As
special cases, we write \(p \oplus q\) for the binary sum and
\(\mathsf{zeroPoly}\) for the nullary sum (the empty interface). An
implementation \(p \Rightarrow \mathsf{Free}\,\mathsf{zeroPoly}\)
depends on no external interfaces---i.e.~it is a \emph{closed}
module---and therefore corresponds precisely to a dependent function
\((a : A) \to B(a)\), since
\(\mathsf{Free}\,\mathsf{zeroPoly}\,C \cong C\).

\section{Program Modules and
Composition}\label{program-modules-and-composition}

\subsection{Example: Fold, Append,
Concat}\label{example-fold-append-concat}

To illustrate the framework, we present a generic fold operation on
lists. The fold module has interface
\(\mathsf{Fold}\,A\,B\,C = (A \times \mathsf{List}\,B,\, \lambda\_ \to C)\)
and depends on two sub-interfaces: a \emph{base case}
\(\mathsf{Base}\,A\,C = (A,\, \lambda\_ \to C)\) and a \emph{recursive
step}
\(\mathsf{Step}\,A\,B\,C = (A \times B \times C,\, \lambda\_ \to C)\).

The implementation proceeds by structural recursion on the input list:
on the empty list, it calls the base case; on a cons cell, it
recursively folds the tail, then calls the step with the accumulated
result:

\begin{Shaded}
\begin{Highlighting}[]
\NormalTok{    fold }\OtherTok{:} \OtherTok{\{}\NormalTok{A B C }\OtherTok{:} \DataTypeTok{Set}\OtherTok{\}}
         \OtherTok{→}\NormalTok{ Fold A B C}
\NormalTok{           ⇒ Free }\OtherTok{(}\NormalTok{sum foldlabels}
                       \OtherTok{(λ\{}\NormalTok{ base }\OtherTok{→}\NormalTok{ Base A C}
                         \OtherTok{;}\NormalTok{ step }\OtherTok{→}\NormalTok{ Step A B C}\OtherTok{\}))}
\NormalTok{    fold }\OtherTok{(}\NormalTok{a , nil}\OtherTok{)} \OtherTok{=}
\NormalTok{        call[ c ← }\OtherTok{(}\NormalTok{base , a}\OtherTok{)}\NormalTok{ ]}
\NormalTok{        return c}
\NormalTok{    fold }\OtherTok{(}\NormalTok{a , cons b bs}\OtherTok{)} \OtherTok{=}
\NormalTok{        do[ c ← fold }\OtherTok{(}\NormalTok{a , bs}\OtherTok{)}\NormalTok{ ]}
\NormalTok{        call[ c\textquotesingle{} ← }\OtherTok{(}\NormalTok{step , }\OtherTok{(}\NormalTok{a , b , c}\OtherTok{))}\NormalTok{ ]}
\NormalTok{        return c\textquotesingle{}}
\end{Highlighting}
\end{Shaded}

Specific instances of folds, such as appending two lists or
concatenating a list of lists, are then obtained by providing closed
implementations of the base and step interfaces and plugging them into
the fold module. For append, the base case returns the accumulator
unchanged, and the step prepends an element---both are closed (depending
on \(\mathsf{zeroPoly}\)). Concat is obtained similarly, using
\texttt{append} as a subroutine for its step case (see Appendix D for
the full code).

\subsection{Composing Implementations}\label{composing-implementations}

Given \(f : p \Rightarrow \mathsf{Free}\,q\) and
\(g : q \Rightarrow \mathsf{Free}\,r\), we can substitute \(g\) for each
call to \(q\) in the computations produced by \(f\), obtaining a
composite \(f \circ g : p \Rightarrow \mathsf{Free}\,r\). This is
precisely the Kleisli composition for the free monad monad on
\(\mathbf{Poly}\): given \(e : \mathsf{Free}\,p\,E\), we recursively
substitute \(g\) for each \texttt{bind} node, using
\texttt{\textgreater{}\textgreater{}=} to splice the resulting trees.
More generally, when a module depends on a \emph{sum} of interfaces, we
can compose it with a family of implementations:

\begin{Shaded}
\begin{Highlighting}[]
\NormalTok{comp }\OtherTok{:} \OtherTok{(}\NormalTok{U }\OtherTok{:} \DataTypeTok{Set}\OtherTok{)} \OtherTok{\{}\NormalTok{p }\OtherTok{:}\NormalTok{ Poly}\OtherTok{\}} \OtherTok{\{}\NormalTok{q }\OtherTok{:}\NormalTok{ U }\OtherTok{→}\NormalTok{ Poly}\OtherTok{\}} \OtherTok{\{}\NormalTok{r }\OtherTok{:}\NormalTok{ Poly}\OtherTok{\}}
     \OtherTok{→} \OtherTok{(}\NormalTok{p ⇒ Free }\OtherTok{(}\NormalTok{sum U q}\OtherTok{))}
     \OtherTok{→} \OtherTok{((}\NormalTok{u }\OtherTok{:}\NormalTok{ U}\OtherTok{)} \OtherTok{→}\NormalTok{ q u ⇒ Free r}\OtherTok{)}
     \OtherTok{→}\NormalTok{ p ⇒ Free r}
\NormalTok{comp U f g }\OtherTok{=}\NormalTok{ f ∘ }\OtherTok{(λ} \OtherTok{(}\NormalTok{u , x}\OtherTok{)} \OtherTok{→}\NormalTok{ g u x}\OtherTok{)}
\end{Highlighting}
\end{Shaded}

Using these combinators, we compose fold with the append base and step
to obtain a closed module implementing append, and similarly for concat,
where the step case composes with append to call it as a subroutine:

\begin{Shaded}
\begin{Highlighting}[]
\NormalTok{    append }\OtherTok{:} \OtherTok{\{}\NormalTok{A }\OtherTok{:} \DataTypeTok{Set}\OtherTok{\}}
           \OtherTok{→}\NormalTok{ Fold }\OtherTok{(}\NormalTok{List A}\OtherTok{)}\NormalTok{ A }\OtherTok{(}\NormalTok{List A}\OtherTok{)}\NormalTok{ ⇒ Free zeroPoly}
\NormalTok{    append }\OtherTok{=}
\NormalTok{        comp foldlabels fold}
             \OtherTok{(λ\{}\NormalTok{ base }\OtherTok{→}\NormalTok{ appendBase}
               \OtherTok{;}\NormalTok{ step }\OtherTok{→}\NormalTok{ appendStep}\OtherTok{\})}

\NormalTok{    concat }\OtherTok{:} \OtherTok{\{}\NormalTok{A }\OtherTok{:} \DataTypeTok{Set}\OtherTok{\}}
           \OtherTok{→}\NormalTok{ Fold ⊤ }\OtherTok{(}\NormalTok{List A}\OtherTok{)} \OtherTok{(}\NormalTok{List A}\OtherTok{)}\NormalTok{ ⇒ Free zeroPoly}
\NormalTok{    concat }\OtherTok{=}
\NormalTok{        comp foldlabels fold}
             \OtherTok{(λ\{}\NormalTok{ base }\OtherTok{→}\NormalTok{ concatBase}
               \OtherTok{;}\NormalTok{ step }\OtherTok{→}\NormalTok{ concatStep ∘ append}\OtherTok{\})}
\end{Highlighting}
\end{Shaded}

\subsection{Wiring Diagrams}\label{wiring-diagrams}

The pattern of composing modules along dependency structures can be
generalized using \emph{wiring diagrams} (Spivak 2013), where boxes
represent modules, wires represent dependencies, and the outer boundary
represents the composite interface. A wiring diagram is parameterized by
a set of \texttt{Boxes} with their \texttt{Arity} (dependencies),
\texttt{Dom}/\texttt{Cod} (dependency and output interfaces), and
\texttt{Inputs} (external dependencies):

\begin{Shaded}
\begin{Highlighting}[]
\KeywordTok{module}\NormalTok{ WD }\OtherTok{(}\NormalTok{Boxes }\OtherTok{:} \DataTypeTok{Set}\OtherTok{)}
          \OtherTok{(}\NormalTok{Arity }\OtherTok{:}\NormalTok{ Boxes }\OtherTok{→} \DataTypeTok{Set}\OtherTok{)}
          \OtherTok{(}\NormalTok{Dom }\OtherTok{:} \OtherTok{(}\NormalTok{b }\OtherTok{:}\NormalTok{ Boxes}\OtherTok{)} \OtherTok{→}\NormalTok{ Arity b }\OtherTok{→}\NormalTok{ Poly}\OtherTok{)}
          \OtherTok{(}\NormalTok{Cod }\OtherTok{:}\NormalTok{ Boxes }\OtherTok{→}\NormalTok{ Poly}\OtherTok{)}
          \OtherTok{(}\NormalTok{Inputs }\OtherTok{:} \DataTypeTok{Set}\OtherTok{)} \OtherTok{(}\NormalTok{inputs }\OtherTok{:}\NormalTok{ Inputs }\OtherTok{→}\NormalTok{ Poly}\OtherTok{)} \KeywordTok{where}

    \KeywordTok{data}\NormalTok{ Wiring }\OtherTok{:} \OtherTok{(}\NormalTok{output }\OtherTok{:}\NormalTok{ Poly}\OtherTok{)} \OtherTok{→} \DataTypeTok{Set₁} \KeywordTok{where}
\NormalTok{        wire }\OtherTok{:} \OtherTok{(}\NormalTok{i }\OtherTok{:}\NormalTok{ Inputs}\OtherTok{)} \OtherTok{→}\NormalTok{ Wiring }\OtherTok{(}\NormalTok{inputs i}\OtherTok{)}
\NormalTok{        box }\OtherTok{:} \OtherTok{(}\NormalTok{b }\OtherTok{:}\NormalTok{ Boxes}\OtherTok{)} \OtherTok{→} \OtherTok{((}\NormalTok{a }\OtherTok{:}\NormalTok{ Arity b}\OtherTok{)} \OtherTok{→}\NormalTok{ Wiring }\OtherTok{(}\NormalTok{Dom b a}\OtherTok{))}
            \OtherTok{→}\NormalTok{ Wiring }\OtherTok{(}\NormalTok{Cod b}\OtherTok{)}
\end{Highlighting}
\end{Shaded}

The constructor \texttt{wire\ i} represents an external dependency
routed in from outside the diagram, while \texttt{box\ b\ f} places box
\texttt{b} in the diagram and wires each of its dependencies
\texttt{a\ :\ Arity\ b} to the sub-diagram \texttt{f\ a}.

The key theorem is that every wiring diagram, together with
implementations for all of its boxes, composes to a single
implementation of the outer interface:

\begin{theorem}\label{thm:compose}
Given a wiring diagram with output interface $p$ and external inputs $\mathsf{inputs}$, together with an implementation of each box $b$ as a morphism $\mathsf{Cod}(b) \Rightarrow \mathsf{Free}(\mathsf{sum}\,(\mathsf{Arity}\,b)\,(\mathsf{Dom}\,b))$, there is an induced implementation $p \Rightarrow \mathsf{Free}(\mathsf{sum}\,\mathsf{Inputs}\,\mathsf{inputs})$.
\end{theorem}

The proof is by induction on the wiring diagram: for a \texttt{wire\ i},
the external input is forwarded; for a \texttt{box\ b\ f}, we compose
the implementation of box \texttt{b} with the recursively composed
sub-diagrams:

\begin{Shaded}
\begin{Highlighting}[]
\NormalTok{compose }\OtherTok{:}\NormalTok{ Wiring Boxes Arity Dom Cod Inputs inputs output}
        \OtherTok{→} \OtherTok{((}\NormalTok{b }\OtherTok{:}\NormalTok{ Boxes}\OtherTok{)} \OtherTok{→}\NormalTok{ Cod b ⇒ Free }\OtherTok{(}\NormalTok{sum }\OtherTok{(}\NormalTok{Arity b}\OtherTok{)} \OtherTok{(}\NormalTok{Dom b}\OtherTok{)))}
        \OtherTok{→}\NormalTok{ output ⇒ Free }\OtherTok{(}\NormalTok{sum Inputs inputs}\OtherTok{)}
\NormalTok{compose }\OtherTok{(}\NormalTok{wire i}\OtherTok{)}\NormalTok{ f x }\OtherTok{=}\NormalTok{ bind }\OtherTok{(}\NormalTok{i , x}\OtherTok{)}\NormalTok{ return}
\NormalTok{compose }\OtherTok{(}\NormalTok{box b f}\OtherTok{)}\NormalTok{ g }\OtherTok{=}\NormalTok{ comp }\OtherTok{\_} \OtherTok{(}\NormalTok{g b}\OtherTok{)} \OtherTok{(λ}\NormalTok{ a }\OtherTok{→}\NormalTok{ compose }\OtherTok{(}\NormalTok{f a}\OtherTok{)}\NormalTok{ g}\OtherTok{)}
\end{Highlighting}
\end{Shaded}

\begin{figure}[ht]
\centering
\begin{tikzpicture}[
    module/.style={draw, rounded corners, minimum width=1.6cm, minimum height=0.7cm, align=center, font=\scriptsize},
    outer/.style={draw, dashed, rounded corners, inner sep=8pt},
    arr/.style={->, >=stealth, semithick},
    label/.style={font=\tiny\itshape}
]
\node[outer, minimum width=5.2cm, minimum height=3.2cm] (outerA) at (-4, 0) {};
\node[above, font=\scriptsize\bfseries] at (outerA.north) {append};
\node[module] (baseA) at (-5, 0.3) {appendBase};
\node[module] (stepA) at (-3, 0.3) {appendStep};
\node[module] (foldA) at (-4, -0.9) {fold};
\draw[arr] (baseA.south) -- node[left, label] {Base} (foldA.north west);
\draw[arr] (stepA.south) -- node[right, label] {Step} (foldA.north east);
\draw[arr] (foldA.south) -- ++(0, -0.3) node[below, label] {out};

\node[outer, minimum width=5.5cm, minimum height=4.2cm] (outerC) at (3, 0) {};
\node[above, font=\scriptsize\bfseries] at (outerC.north) {concat};
\node[module] (baseC) at (1.5, 0.1) {concatBase};
\node[module] (stepC) at (4.5, 0.1) {concatStep};
\node[module] (appendC) at (4.5, 1.5) {append};
\node[module] (foldC) at (3.0, -1.4) {fold};
\draw[arr] (baseC.south) -- node[left, label] {Base} (foldC.north west);
\draw[arr] (stepC.south) -- node[right, label] {Step} (foldC.north east);
\draw[arr] (appendC.south) -- node[right, label] {Fold} (stepC.north);
\draw[arr] (foldC.south) -- ++(0, -0.3) node[below, label] {out};
\end{tikzpicture}
\caption{Wiring diagrams for \texttt{append} (left) and \texttt{concat} (right). In \texttt{concat}, the step box's dependency is wired through the \texttt{append} module.}
\label{fig:wiring-diagrams}
\end{figure}
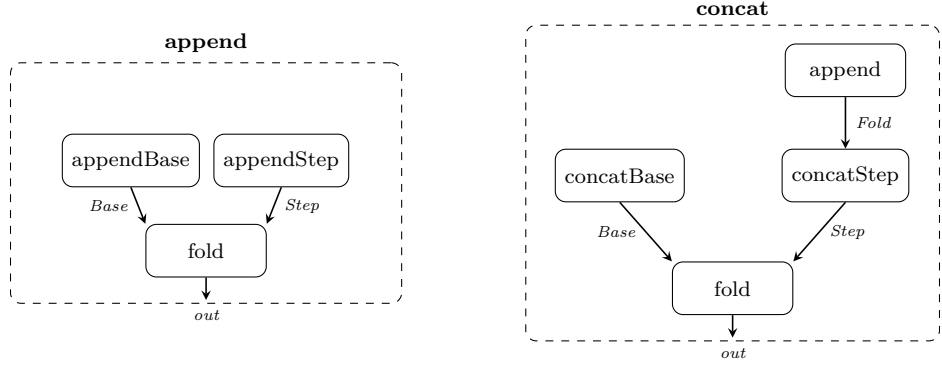

As a concrete illustration, Figure \ref{fig:wiring-diagrams} shows the
wiring diagrams for \texttt{append} and \texttt{concat}.

\section{Coalgebraic Operational
Semantics}\label{coalgebraic-operational-semantics}

\subsection{Mealy Machines}\label{mealy-machines}

To actually \emph{run} programs built up in the manner described above,
we need an \emph{operational semantics}. For this purpose, we develop a
\emph{coalgebraic} approach based on \emph{Mealy machines} (Niu and
Spivak 2024, Ch. 4; Myers 2022):

\begin{definition}\label{def:mealy}
A \emph{Mealy machine} with interface $p = (A, B)$ is an element of the following coinductive record type:
\end{definition}

\begin{Shaded}
\begin{Highlighting}[]
\KeywordTok{record}\NormalTok{ Mealy }\OtherTok{(}\NormalTok{p }\OtherTok{:}\NormalTok{ Poly}\OtherTok{)} \OtherTok{:} \DataTypeTok{Set} \KeywordTok{where}
    \KeywordTok{coinductive}
    \KeywordTok{field}
\NormalTok{        app }\OtherTok{:} \OtherTok{(}\NormalTok{x }\OtherTok{:}\NormalTok{ p }\OtherTok{.}\NormalTok{fst}\OtherTok{)} \OtherTok{→}\NormalTok{ p }\OtherTok{.}\NormalTok{snd x × Mealy p}
\end{Highlighting}
\end{Shaded}

In other words, a Mealy machine with interface \((A, B)\) is a stateful
automaton that, given an input \(a : A\), produces an output of type
\(B(a)\) together with an updated machine state. The type
\(\mathsf{Mealy}\,p\) is equivalently the \emph{final coalgebra} of the
functor \(X \mapsto \prod_{a:A} B(a) \times X\).

Given a Mealy machine with interface \(p\), we can \emph{run} a
computation \(e : \mathsf{Free}\,p\,C\) on it by threading the machine
state through each call, yielding a result and an updated machine. More
generally, given an implementation
\(f : p \Rightarrow \mathsf{Free}\,q\) and a Mealy machine for \(q\), we
obtain a Mealy machine for \(p\):

\begin{Shaded}
\begin{Highlighting}[]
\NormalTok{prog→mealy }\OtherTok{:} \OtherTok{\{}\NormalTok{p q }\OtherTok{:}\NormalTok{ Poly}\OtherTok{\}}
           \OtherTok{→} \OtherTok{(}\NormalTok{f }\OtherTok{:}\NormalTok{ p ⇒ Free q}\OtherTok{)} \OtherTok{→}\NormalTok{ Mealy q }\OtherTok{→}\NormalTok{ Mealy p}
\NormalTok{prog→mealy f m }\OtherTok{.}\NormalTok{app a }\OtherTok{=}
    \KeywordTok{let} \OtherTok{(}\NormalTok{b , m\textquotesingle{}}\OtherTok{)} \OtherTok{=}\NormalTok{ run{-}mealy }\OtherTok{(}\NormalTok{f a}\OtherTok{)}\NormalTok{ m }\KeywordTok{in}
    \OtherTok{(}\NormalTok{b , prog→mealy f m\textquotesingle{}}\OtherTok{)}
\end{Highlighting}
\end{Shaded}

Moreover, Mealy machines are closed under the coproduct of polynomials,
so a pair of Mealy machines for \(p\) and \(q\) can be combined into one
for \(p \oplus q\), as follows:

\begin{Shaded}
\begin{Highlighting}[]
\OtherTok{\_}\NormalTok{mealy⊕}\OtherTok{\_} \OtherTok{:} \OtherTok{\{}\NormalTok{p q }\OtherTok{:}\NormalTok{ Poly}\OtherTok{\}} 
         \OtherTok{→}\NormalTok{ Mealy p }\OtherTok{→}\NormalTok{ Mealy q }\OtherTok{→}\NormalTok{ Mealy }\OtherTok{(}\NormalTok{p ⊕ q}\OtherTok{)}
\OtherTok{\_}\NormalTok{mealy⊕}\OtherTok{\_}\NormalTok{ m₁ m₂ }\OtherTok{.}\NormalTok{app }\OtherTok{(}\NormalTok{inl a}\OtherTok{)} \OtherTok{=} 
    \KeywordTok{let} \OtherTok{(}\NormalTok{b , m\textquotesingle{}}\OtherTok{)} \OtherTok{=}\NormalTok{ m₁ }\OtherTok{.}\NormalTok{app a }\KeywordTok{in} \OtherTok{(}\NormalTok{b , }\OtherTok{(}\NormalTok{m\textquotesingle{} mealy⊕ m₂}\OtherTok{))}
\OtherTok{\_}\NormalTok{mealy⊕}\OtherTok{\_}\NormalTok{ m₁ m₂ }\OtherTok{.}\NormalTok{app }\OtherTok{(}\NormalTok{inr c}\OtherTok{)} \OtherTok{=} 
    \KeywordTok{let} \OtherTok{(}\NormalTok{d , m\textquotesingle{}}\OtherTok{)} \OtherTok{=}\NormalTok{ m₂ }\OtherTok{.}\NormalTok{app c }\KeywordTok{in} \OtherTok{(}\NormalTok{d , }\OtherTok{(}\NormalTok{m₁ mealy⊕ m\textquotesingle{}}\OtherTok{))}
\end{Highlighting}
\end{Shaded}

Together, \texttt{prog→mealy} and \texttt{\_mealy⊕\_} give a functorial
operational semantics: given Mealy machines \(m_1, m_2\) with interfaces
\(p\) and \(q\), and a program module
\(f : r \Rightarrow \mathsf{Free}\,(p \oplus q)\) specifying a pattern
of interaction between them, we obtain a Mealy machine for the interface
\(r\) that executes the specified interactions. This approach is closely
related to the module structure of the free monad monad over the cofree
comonad comonad described in Libkind and Spivak's ``Pattern Runs on
Matter'' (Libkind and Spivak 2025). Details are given in Appendix B.

\subsection{Effects and Runners}\label{effects-and-runners}

The fact that Mealy machines represent \emph{stateful} computations, yet
can be composed along program modules, means that the functions
specified by a polynomial interface need not be \emph{pure}; in
particular, they can have \emph{effects} that are implemented by the
Mealy machine.

A polynomial morphism \(f : p \Rightarrow \mathsf{Free}\,q\) can thus be
seen as using \emph{effects} encoded by \(q\), and a Mealy machine for
\(q\) then corresponds to a \emph{runner} for these effects, as defined
by Ahman and Bauer (Ahman and Bauer 2020). As a concrete example, the
\emph{state effect} for a type \(S\) is encoded by the following
interface:

\begin{Shaded}
\begin{Highlighting}[]
\NormalTok{State }\OtherTok{:} \DataTypeTok{Set} \OtherTok{→}\NormalTok{ Poly}
\NormalTok{State S }\OtherTok{.}\NormalTok{fst }\OtherTok{=}\NormalTok{ StateI S}
\NormalTok{State S }\OtherTok{.}\NormalTok{snd get }\OtherTok{=}\NormalTok{ S}
\NormalTok{State S }\OtherTok{.}\NormalTok{snd }\OtherTok{(}\NormalTok{put s}\OtherTok{)} \OtherTok{=}\NormalTok{ ⊤}
\end{Highlighting}
\end{Shaded}

where \(\mathsf{StateI}\,S\) has constructors \(\mathsf{get}\) and
\(\mathsf{put}\,s\) for \(s : S\), corresponding to operations for
reading from and writing to the state, respectively. A Mealy machine
implementing this interface simply threads the current state:

\begin{Shaded}
\begin{Highlighting}[]
\NormalTok{state{-}mealy }\OtherTok{:} \OtherTok{\{}\NormalTok{S }\OtherTok{:} \DataTypeTok{Set}\OtherTok{\}} \OtherTok{→}\NormalTok{ S }\OtherTok{→}\NormalTok{ Mealy }\OtherTok{(}\NormalTok{State S}\OtherTok{)}
\NormalTok{state{-}mealy s }\OtherTok{.}\NormalTok{app get }\OtherTok{=} \OtherTok{(}\NormalTok{s , state{-}mealy s}\OtherTok{)}
\NormalTok{state{-}mealy s }\OtherTok{.}\NormalTok{app }\OtherTok{(}\NormalTok{put s\textquotesingle{}}\OtherTok{)} \OtherTok{=} \OtherTok{(\_}\NormalTok{ , state{-}mealy s\textquotesingle{}}\OtherTok{)}
\end{Highlighting}
\end{Shaded}

\subsection{Example: Fibonacci}\label{example-fibonacci}

Using the state effect, we implement a Fibonacci Mealy machine. The
update reads the current pair of Fibonacci numbers from state, advances
it, and returns the first component:

\begin{Shaded}
\begin{Highlighting}[]
\NormalTok{    fib{-}update }\OtherTok{:} \OtherTok{(}\NormalTok{⊤ , }\OtherTok{λ} \OtherTok{\_} \OtherTok{→}\NormalTok{ Nat}\OtherTok{)}\NormalTok{ ⇒ Free }\OtherTok{(}\NormalTok{State }\OtherTok{(}\NormalTok{Nat × Nat}\OtherTok{))}
\NormalTok{    fib{-}update }\OtherTok{\_} \OtherTok{=}
\NormalTok{        call[ }\OtherTok{(}\NormalTok{x , y}\OtherTok{)}\NormalTok{ ← get ]}
\NormalTok{        call[ }\OtherTok{\_}\NormalTok{ ← put }\OtherTok{(}\NormalTok{y , x + y}\OtherTok{)}\NormalTok{ ]}
\NormalTok{        return x}

\NormalTok{    fib }\OtherTok{:}\NormalTok{ Mealy }\OtherTok{(}\NormalTok{⊤ , }\OtherTok{λ} \OtherTok{\_} \OtherTok{→}\NormalTok{ Nat}\OtherTok{)}
\NormalTok{    fib }\OtherTok{=}\NormalTok{ prog→mealy fib{-}update }\OtherTok{(}\NormalTok{state{-}mealy }\OtherTok{(}\DecValTok{0}\NormalTok{ , }\DecValTok{1}\OtherTok{))}
\end{Highlighting}
\end{Shaded}

The machine \texttt{fib} is obtained by composing the update program
with the state runner initialized to \((0, 1)\). On each invocation, it
outputs \(0, 1, 1, 2, 3, 5, \ldots\) in sequence.

\section{Dependent Polynomials and Compositional
Verification}\label{dependent-polynomials-and-compositional-verification}

We now come to the heart of the paper: representing
\emph{specifications} of program modules in a way that supports
compositional verification.

\subsection{Dependent Polynomials as
Specifications}\label{dependent-polynomials-as-specifications}

\begin{definition}\label{def:deppoly}
A \emph{dependent polynomial} over a polynomial $p = (A, B)$ consists of:
\begin{itemize}
    \item a family of \emph{preconditions} $C : A \to \mathsf{Set}$ on inputs $a : A$, and
    \item a family of \emph{postconditions} $D : (a : A) \to C(a) \to B(a) \to \mathsf{Set}$ on outputs $b : B(a)$, depending on $a : A$ and evidence $c : C(a)$ that the precondition holds.
\end{itemize}
\end{definition}

\begin{Shaded}
\begin{Highlighting}[]
\NormalTok{DepPoly }\OtherTok{:}\NormalTok{ Poly }\OtherTok{→} \DataTypeTok{Set₁}
\NormalTok{DepPoly }\OtherTok{(}\NormalTok{A , B}\OtherTok{)} \OtherTok{=}\NormalTok{ Σ }\OtherTok{(}\NormalTok{A }\OtherTok{→} \DataTypeTok{Set}\OtherTok{)} \OtherTok{(λ}\NormalTok{ C }\OtherTok{→} \OtherTok{(}\NormalTok{x }\OtherTok{:}\NormalTok{ A}\OtherTok{)} \OtherTok{→}\NormalTok{ C x }\OtherTok{→}\NormalTok{ B x }\OtherTok{→} \DataTypeTok{Set}\OtherTok{)}
\end{Highlighting}
\end{Shaded}

If \(p = (A, B)\) represents the interface of a function
\(f : (a : A) \to B(a)\), then a dependent polynomial \((C, D)\) over
\(p\) specifies a \emph{contract}: for each input \(a\) satisfying
precondition \(C(a)\), the output \(f(a)\) should satisfy postcondition
\(D(a, c, f(a))\), where \(c\) is the evidence that the precondition
holds. This is the polynomial-functorial analogue of a \emph{Hoare
triple} \(\{C(a)\}\, f(a) \,\{D(a, -, -)\}\).

Categorically, a dependent polynomial over \(p\) is equivalently a
polynomial functor on the arrow category \(\mathbf{Type}^{\to}\) lying
over \(p\) via the codomain functor
\(\mathsf{cod} : \mathbf{Type}^{\to} \to \mathbf{Type}\).

A \emph{dependent morphism} \(\mathsf{\Rightarrow Dep}\,p\,r\,G\,f\)
over a morphism \(f : p \Rightarrow F\)---where \(F\) is an endofunctor
on types and \(G\) is a \emph{dependent} endofunctor, i.e.~an
endofunctor on \(\mathbf{Type}^\to\) lying over \(F\) via the codomain
map---is a function that, for each input \(a\) and evidence \(c : C(a)\)
that the precondition holds, produces a \(G\)-structure witnessing that
the postcondition will be satisfied on the output \(f(a)\).

\begin{Shaded}
\begin{Highlighting}[]
\NormalTok{⇒Dep }\OtherTok{:} \OtherTok{(}\NormalTok{p }\OtherTok{:}\NormalTok{ Poly}\OtherTok{)} \OtherTok{→}\NormalTok{ DepPoly p }\OtherTok{→} \OtherTok{\{}\NormalTok{F }\OtherTok{:} \DataTypeTok{Set} \OtherTok{→} \DataTypeTok{Set}\OtherTok{\}}
     \OtherTok{→} \OtherTok{((}\NormalTok{X }\OtherTok{:} \DataTypeTok{Set}\OtherTok{)} \OtherTok{→} \OtherTok{(}\NormalTok{X }\OtherTok{→} \DataTypeTok{Set}\OtherTok{)} \OtherTok{→}\NormalTok{ F X }\OtherTok{→} \DataTypeTok{Set}\OtherTok{)}
     \OtherTok{→} \OtherTok{(}\NormalTok{f }\OtherTok{:}\NormalTok{ p ⇒ F}\OtherTok{)} \OtherTok{→} \DataTypeTok{Set}
\NormalTok{⇒Dep }\OtherTok{(}\NormalTok{A , B}\OtherTok{)} \OtherTok{(}\NormalTok{C , D}\OtherTok{)}\NormalTok{ G f }\OtherTok{=}
    \OtherTok{(}\NormalTok{a }\OtherTok{:}\NormalTok{ A}\OtherTok{)} \OtherTok{(}\NormalTok{c }\OtherTok{:}\NormalTok{ C a}\OtherTok{)} \OtherTok{→}\NormalTok{ G }\OtherTok{(}\NormalTok{B a}\OtherTok{)} \OtherTok{(}\NormalTok{D a c}\OtherTok{)} \OtherTok{(}\NormalTok{f a}\OtherTok{)}
\end{Highlighting}
\end{Shaded}

\subsection{Dependent Free Monads}\label{dependent-free-monads}

To verify implementations \(f : p \Rightarrow \mathsf{Free}\,q\), we
need a notion of \emph{verified computation} over the free monad. This
is provided by the \emph{dependent free monad}, which is equivalently
the free monad monad on the category of dependent polynomials:

\begin{Shaded}
\begin{Highlighting}[]
\KeywordTok{data}\NormalTok{ FreeDep }\OtherTok{(}\NormalTok{p }\OtherTok{:}\NormalTok{ Poly}\OtherTok{)} \OtherTok{(}\NormalTok{r }\OtherTok{:}\NormalTok{ DepPoly p}\OtherTok{)}
             \OtherTok{(}\NormalTok{E }\OtherTok{:} \DataTypeTok{Set}\OtherTok{)} \OtherTok{(}\NormalTok{F }\OtherTok{:}\NormalTok{ E }\OtherTok{→} \DataTypeTok{Set}\OtherTok{)} \OtherTok{:}\NormalTok{ Free p E }\OtherTok{→} \DataTypeTok{Set} \KeywordTok{where}
\NormalTok{    returnD }\OtherTok{:} \OtherTok{\{}\NormalTok{e }\OtherTok{:}\NormalTok{ E}\OtherTok{\}} \OtherTok{→}\NormalTok{ F e }\OtherTok{→}\NormalTok{ FreeDep p r E F }\OtherTok{(}\NormalTok{return e}\OtherTok{)}
\NormalTok{    bindD }\OtherTok{:} \OtherTok{\{}\NormalTok{a }\OtherTok{:}\NormalTok{ p }\OtherTok{.}\NormalTok{fst}\OtherTok{\}} \OtherTok{(}\NormalTok{c }\OtherTok{:}\NormalTok{ r }\OtherTok{.}\NormalTok{fst a}\OtherTok{)}
          \OtherTok{→} \OtherTok{\{}\NormalTok{k }\OtherTok{:} \OtherTok{(}\NormalTok{p }\OtherTok{.}\NormalTok{snd a}\OtherTok{)} \OtherTok{→}\NormalTok{ Free p E}\OtherTok{\}}
          \OtherTok{→} \OtherTok{((}\NormalTok{b }\OtherTok{:}\NormalTok{ p }\OtherTok{.}\NormalTok{snd a}\OtherTok{)} \OtherTok{→}\NormalTok{ r }\OtherTok{.}\NormalTok{snd a c b}
                            \OtherTok{→}\NormalTok{ FreeDep p r E F }\OtherTok{(}\NormalTok{k b}\OtherTok{))}
          \OtherTok{→}\NormalTok{ FreeDep p r E F }\OtherTok{(}\NormalTok{bind a k}\OtherTok{)}
\end{Highlighting}
\end{Shaded}

An element of \(\mathsf{FreeDep}\,p\,r\,E\,F\,e\) proves that
computation \(e\) respects specification \(r\): assuming each call's
response satisfies the postcondition, the final result satisfies
\(F\)---this is \emph{assume-guarantee} reasoning now lifted to the
realm of dependent types.

\begin{definition}\label{def:verified-impl}
A \emph{verified implementation} of specification $(p, r)$ depending on specification $(q, s)$ consists of an implementation $f : p \Rightarrow \mathsf{Free}\,q$ together with a dependent morphism of type
$$\mathsf{\Rightarrow Dep}\,p\,r\,(\mathsf{FreeDep}\,q\,s)\,f$$
witnessing that $f$ satisfies specification $r$ assuming $q$ satisfies specification $s$.
\end{definition}

The crucial compositionality theorem is then that verified
implementations compose in the same manner as implementations:

\begin{theorem}\label{thm:comp-dep}
Given a wiring diagram with implementations for each box, and specifications for the interfaces of each box and of the outer box as a whole, along with verifications for each box's specification, the composite implementation carries a composite verification for the specifications of the outer interfaces. That is, verifications compose along wiring diagrams in the same way that their underlying implementations do.
\end{theorem}

The proof uses the dependent analogue \texttt{∘Dep} of Kleisli
composition for the dependent free monad monad, along with the fact that
dependent polynomials are closed under sums:

\begin{Shaded}
\begin{Highlighting}[]
\CommentTok{{-}{-} sums of dependent polynomials}
\NormalTok{sumDep }\OtherTok{:} \OtherTok{(}\NormalTok{U }\OtherTok{:} \DataTypeTok{Set}\OtherTok{)} \OtherTok{(}\NormalTok{p }\OtherTok{:}\NormalTok{ U }\OtherTok{→}\NormalTok{ Poly}\OtherTok{)} 
       \OtherTok{→} \OtherTok{(}\NormalTok{r }\OtherTok{:} \OtherTok{(}\NormalTok{u }\OtherTok{:}\NormalTok{ U}\OtherTok{)} \OtherTok{→}\NormalTok{ DepPoly }\OtherTok{(}\NormalTok{p u}\OtherTok{))} \OtherTok{→}\NormalTok{ DepPoly }\OtherTok{(}\NormalTok{sum U p}\OtherTok{)}
\NormalTok{sumDep U p r }\OtherTok{.}\NormalTok{fst }\OtherTok{(}\NormalTok{u , x}\OtherTok{)} \OtherTok{=}\NormalTok{ r u }\OtherTok{.}\NormalTok{fst x}
\NormalTok{sumDep U p r }\OtherTok{.}\NormalTok{snd }\OtherTok{(}\NormalTok{u , x}\OtherTok{)}\NormalTok{ y }\OtherTok{=}\NormalTok{ r u }\OtherTok{.}\NormalTok{snd x y}
\end{Highlighting}
\end{Shaded}

The multi-ary composition \texttt{compDep} then mirrors \texttt{comp}.

\begin{Shaded}
\begin{Highlighting}[]
\NormalTok{compDep }\OtherTok{:} \OtherTok{(}\NormalTok{U }\OtherTok{:} \DataTypeTok{Set}\OtherTok{)} \OtherTok{\{}\NormalTok{p }\OtherTok{:}\NormalTok{ Poly}\OtherTok{\}} \OtherTok{\{}\NormalTok{q }\OtherTok{:}\NormalTok{ U }\OtherTok{→}\NormalTok{ Poly}\OtherTok{\}} \OtherTok{\{}\NormalTok{r }\OtherTok{:}\NormalTok{ Poly}\OtherTok{\}}
          \OtherTok{\{}\NormalTok{s }\OtherTok{:}\NormalTok{ DepPoly p}\OtherTok{\}} \OtherTok{\{}\NormalTok{t }\OtherTok{:} \OtherTok{(}\NormalTok{u }\OtherTok{:}\NormalTok{ U}\OtherTok{)} \OtherTok{→}\NormalTok{ DepPoly }\OtherTok{(}\NormalTok{q u}\OtherTok{)\}}
          \OtherTok{\{}\NormalTok{v }\OtherTok{:}\NormalTok{ DepPoly r}\OtherTok{\}} \OtherTok{\{}\NormalTok{f }\OtherTok{:}\NormalTok{ p ⇒ Free }\OtherTok{(}\NormalTok{sum U q}\OtherTok{)\}}
        \OtherTok{→} \OtherTok{\{}\NormalTok{g }\OtherTok{:} \OtherTok{(}\NormalTok{u }\OtherTok{:}\NormalTok{ U}\OtherTok{)} \OtherTok{→}\NormalTok{ q u ⇒ Free r}\OtherTok{\}}
        \OtherTok{→} \OtherTok{(}\NormalTok{⇒Dep p s }\OtherTok{(}\NormalTok{FreeDep }\OtherTok{(}\NormalTok{sum U q}\OtherTok{)} \OtherTok{(}\NormalTok{sumDep U q t}\OtherTok{))}\NormalTok{ f}\OtherTok{)}
        \OtherTok{→} \OtherTok{((}\NormalTok{u }\OtherTok{:}\NormalTok{ U}\OtherTok{)} \OtherTok{→}\NormalTok{ ⇒Dep }\OtherTok{(}\NormalTok{q u}\OtherTok{)} \OtherTok{(}\NormalTok{t u}\OtherTok{)} \OtherTok{(}\NormalTok{FreeDep r v}\OtherTok{)} \OtherTok{(}\NormalTok{g u}\OtherTok{))}
        \OtherTok{→}\NormalTok{ ⇒Dep p s }\OtherTok{(}\NormalTok{FreeDep r v}\OtherTok{)} \OtherTok{(}\NormalTok{comp U f g}\OtherTok{)}
\NormalTok{compDep U ff gg }\OtherTok{=}\NormalTok{ ∘Dep ff }\OtherTok{(λ} \OtherTok{(}\NormalTok{u , x}\OtherTok{)}\NormalTok{ c }\OtherTok{→}\NormalTok{ gg u x c}\OtherTok{)}
\end{Highlighting}
\end{Shaded}

As an example (Appendix C), we verify that our \texttt{append}
implementation correctly computes list-append. The key idea is that
proofs by \emph{induction}---or more specifically using \emph{loop
invariants}---on recursive functions on lists are simply the dependent
analogue of folds on lists: we define a generic verification
\texttt{foldInd} over the \texttt{fold} module parameterized by
specifications for the base and step. To verify \texttt{append}, we
instantiate this module with a relational specification, and use
\texttt{compDep} to compose the base and step verifications with
\texttt{foldInd}, yielding a complete verification of the append module.

\section{Verified Operational
Semantics}\label{verified-operational-semantics}

\subsection{Dependent Mealy Machines}\label{dependent-mealy-machines}

Just as Mealy machines provide an operational semantics for program
modules, \emph{dependent Mealy machines} provide a notion of
\emph{specification/verification} for these operational semantics:

\begin{definition}\label{def:depmealy}
A \emph{dependent Mealy machine} for specification $(p, r)$ over a Mealy machine $m$ for $p$ is a coinductive record:
\end{definition}

\begin{Shaded}
\begin{Highlighting}[]
\KeywordTok{record}\NormalTok{ DepMealy }\OtherTok{(}\NormalTok{p }\OtherTok{:}\NormalTok{ Poly}\OtherTok{)} \OtherTok{(}\NormalTok{r }\OtherTok{:}\NormalTok{ DepPoly p}\OtherTok{)} \OtherTok{(}\NormalTok{m }\OtherTok{:}\NormalTok{ Mealy p}\OtherTok{)} \OtherTok{:} \DataTypeTok{Set} \KeywordTok{where}
    \KeywordTok{coinductive}
    \KeywordTok{field}
\NormalTok{        appD }\OtherTok{:} \OtherTok{(}\NormalTok{x }\OtherTok{:}\NormalTok{ p }\OtherTok{.}\NormalTok{fst}\OtherTok{)} \OtherTok{(}\NormalTok{y }\OtherTok{:}\NormalTok{ r }\OtherTok{.}\NormalTok{fst x}\OtherTok{)}
             \OtherTok{→}\NormalTok{ r }\OtherTok{.}\NormalTok{snd x y }\OtherTok{(}\NormalTok{m }\OtherTok{.}\NormalTok{app x }\OtherTok{.}\NormalTok{fst}\OtherTok{)}
\NormalTok{               × DepMealy p r }\OtherTok{(}\NormalTok{m }\OtherTok{.}\NormalTok{app x }\OtherTok{.}\NormalTok{snd}\OtherTok{)}
\end{Highlighting}
\end{Shaded}

The key compositionality results carry over to the dependent setting:

\begin{proposition}\label{prop:prog-mealy-dep}
Given a verified implementation $f : p ⇒ Free q$ with respect to specifications $r$ on $p$ and $s$ on $q$, and a dependent Mealy machine for $(q, s)$ over $m$, the induced Mealy machine $\mathsf{prog\to mealy}\,f\,m$ carries a dependent Mealy machine for $(p, r)$.
\end{proposition}

The proof uses \texttt{run-mealyD}, which threads verification evidence
alongside the machine state:

\begin{Shaded}
\begin{Highlighting}[]
\NormalTok{prog→mealyD }\OtherTok{:} \OtherTok{\{}\NormalTok{p q }\OtherTok{:}\NormalTok{ Poly}\OtherTok{\}} \OtherTok{\{}\NormalTok{r }\OtherTok{:}\NormalTok{ DepPoly p}\OtherTok{\}} \OtherTok{\{}\NormalTok{s }\OtherTok{:}\NormalTok{ DepPoly q}\OtherTok{\}}
            \OtherTok{→} \OtherTok{\{}\NormalTok{f }\OtherTok{:}\NormalTok{ p ⇒ Free q}\OtherTok{\}} \OtherTok{\{}\NormalTok{m }\OtherTok{:}\NormalTok{ Mealy q}\OtherTok{\}}
            \OtherTok{→} \OtherTok{(}\NormalTok{⇒Dep p r }\OtherTok{(}\NormalTok{FreeDep q s}\OtherTok{)}\NormalTok{ f}\OtherTok{)} \OtherTok{→}\NormalTok{ DepMealy q s m}
            \OtherTok{→}\NormalTok{ DepMealy p r }\OtherTok{(}\NormalTok{prog→mealy f m}\OtherTok{)}
\NormalTok{prog→mealyD ff mm }\OtherTok{.}\NormalTok{appD a c }\OtherTok{=}
    \KeywordTok{let} \OtherTok{(}\NormalTok{b , mm\textquotesingle{}}\OtherTok{)} \OtherTok{=}\NormalTok{ run{-}mealyD }\OtherTok{(}\NormalTok{ff a c}\OtherTok{)}\NormalTok{ mm }\KeywordTok{in}
    \OtherTok{(}\NormalTok{b , }\OtherTok{(}\NormalTok{prog→mealyD ff mm\textquotesingle{}}\OtherTok{))}
\end{Highlighting}
\end{Shaded}

\begin{proposition}\label{prop:depmealy-sum}
Dependent Mealy machines are closed under binary sums: given a dependent Mealy machine for $(p, r)$ over $m_1$ and one for $(q, s)$ over $m_2$, there is a dependent Mealy machine for $(p \oplus q,\, r \oplus s)$ over $m_1 \oplus m_2$.
\end{proposition}

\begin{Shaded}
\begin{Highlighting}[]
\OtherTok{\_}\NormalTok{depMealy⊕}\OtherTok{\_} \OtherTok{:} \OtherTok{\{}\NormalTok{p q }\OtherTok{:}\NormalTok{ Poly}\OtherTok{\}} \OtherTok{\{}\NormalTok{r }\OtherTok{:}\NormalTok{ DepPoly p}\OtherTok{\}} \OtherTok{\{}\NormalTok{s }\OtherTok{:}\NormalTok{ DepPoly q}\OtherTok{\}}
            \OtherTok{→} \OtherTok{\{}\NormalTok{m₁ }\OtherTok{:}\NormalTok{ Mealy p}\OtherTok{\}} \OtherTok{\{}\NormalTok{m₂ }\OtherTok{:}\NormalTok{ Mealy q}\OtherTok{\}}
            \OtherTok{→}\NormalTok{ DepMealy p r m₁ }\OtherTok{→}\NormalTok{ DepMealy q s m₂}
            \OtherTok{→}\NormalTok{ DepMealy }\OtherTok{(}\NormalTok{p ⊕ q}\OtherTok{)} \OtherTok{(}\NormalTok{depPoly⊕ r s}\OtherTok{)} \OtherTok{(}\NormalTok{m₁ mealy⊕ m₂}\OtherTok{)}
\OtherTok{\_}\NormalTok{depMealy⊕}\OtherTok{\_}\NormalTok{ m₁ m₂ }\OtherTok{.}\NormalTok{appD }\OtherTok{(}\NormalTok{inl a}\OtherTok{)}\NormalTok{ c }\OtherTok{=}
    \KeywordTok{let} \OtherTok{(}\NormalTok{b , m\textquotesingle{}}\OtherTok{)} \OtherTok{=}\NormalTok{ m₁ }\OtherTok{.}\NormalTok{appD a c }\KeywordTok{in}
    \OtherTok{(}\NormalTok{b , }\OtherTok{(}\NormalTok{m\textquotesingle{} depMealy⊕ m₂}\OtherTok{))}
\OtherTok{\_}\NormalTok{depMealy⊕}\OtherTok{\_}\NormalTok{ m₁ m₂ }\OtherTok{.}\NormalTok{appD }\OtherTok{(}\NormalTok{inr c}\OtherTok{)}\NormalTok{ d }\OtherTok{=}
    \KeywordTok{let} \OtherTok{(}\NormalTok{e , m\textquotesingle{}}\OtherTok{)} \OtherTok{=}\NormalTok{ m₂ }\OtherTok{.}\NormalTok{appD c d }\KeywordTok{in}
    \OtherTok{(}\NormalTok{e , }\OtherTok{(}\NormalTok{m₁ depMealy⊕ m\textquotesingle{}}\OtherTok{))}
\end{Highlighting}
\end{Shaded}

\subsection{State Invariants}\label{state-invariants}

A natural application is verifying \emph{state invariants}. Given
\(T : S \to \mathsf{Set}\), we specify that if the initial state
satisfies \(T\) and all written states satisfy \(T\), then all read
states satisfy \(T\):

\begin{Shaded}
\begin{Highlighting}[]
\CommentTok{{-}{-} specification for state invariants}
\NormalTok{Invariant }\OtherTok{:} \OtherTok{(}\NormalTok{S }\OtherTok{:} \DataTypeTok{Set}\OtherTok{)} \OtherTok{(}\NormalTok{T }\OtherTok{:}\NormalTok{ S }\OtherTok{→} \DataTypeTok{Set}\OtherTok{)} \OtherTok{→}\NormalTok{ DepPoly }\OtherTok{(}\NormalTok{State S}\OtherTok{)}
\NormalTok{Invariant S T }\OtherTok{.}\NormalTok{fst get }\OtherTok{=}\NormalTok{ ⊤}
\NormalTok{Invariant S T }\OtherTok{.}\NormalTok{fst }\OtherTok{(}\NormalTok{put s}\OtherTok{)} \OtherTok{=}\NormalTok{ T s}
\NormalTok{Invariant S T }\OtherTok{.}\NormalTok{snd get }\OtherTok{\_}\NormalTok{ s }\OtherTok{=}\NormalTok{ T s}
\NormalTok{Invariant S T }\OtherTok{.}\NormalTok{snd }\OtherTok{(}\NormalTok{put s}\OtherTok{)}\NormalTok{ t }\OtherTok{\_} \OtherTok{=}\NormalTok{ ⊤}

\CommentTok{{-}{-} verification of a state invariant}
\NormalTok{invariant }\OtherTok{:} \OtherTok{\{}\NormalTok{S }\OtherTok{:} \DataTypeTok{Set}\OtherTok{\}} \OtherTok{\{}\NormalTok{T }\OtherTok{:}\NormalTok{ S }\OtherTok{→} \DataTypeTok{Set}\OtherTok{\}}
          \OtherTok{→} \OtherTok{(}\NormalTok{s }\OtherTok{:}\NormalTok{ S}\OtherTok{)} \OtherTok{→}\NormalTok{ T s }
          \OtherTok{→}\NormalTok{ DepMealy }\OtherTok{(}\NormalTok{State S}\OtherTok{)} \OtherTok{(}\NormalTok{Invariant S T}\OtherTok{)} \OtherTok{(}\NormalTok{state{-}mealy s}\OtherTok{)}
\NormalTok{invariant s t }\OtherTok{.}\NormalTok{appD get }\OtherTok{\_} \OtherTok{=} \OtherTok{(}\NormalTok{t , }\OtherTok{(}\NormalTok{invariant s t}\OtherTok{))}
\NormalTok{invariant s t }\OtherTok{.}\NormalTok{appD }\OtherTok{(}\NormalTok{put s\textquotesingle{}}\OtherTok{)}\NormalTok{ t\textquotesingle{} }\OtherTok{=} \OtherTok{(\_}\NormalTok{ , invariant s\textquotesingle{} t\textquotesingle{}}\OtherTok{)}
\end{Highlighting}
\end{Shaded}

The verification is a straightforward coinductive argument.

\subsection{Example: Verified
Fibonacci}\label{example-verified-fibonacci}

We can verify that the \texttt{fib} Mealy machine computes Fibonacci
numbers by defining a relational specification \texttt{IsFib} and a
state invariant asserting that the state \((y, z)\) satisfies
\(\mathsf{IsFib}(x, y) \times \mathsf{IsFib}(x+1, z)\) for some index
\(x\). The verification of \texttt{fib-update} shows that each update
step preserves this invariant and returns the correct Fibonacci number.
Composing with the state invariant verification via \texttt{prog→mealyD}
yields a full dependent Mealy machine for the specification. The full
code is given in Appendix C.

\section{Categorical Perspective}\label{categorical-perspective}

Stepping back from the concrete constructions, we can identify the
abstract categorical structure that makes the above-described
compositional verification work in general.

Let $\mathbf{Int}$ (the category of \emph{interfaces}) be the category whose objects are polynomial functors and whose morphisms $p \to q$ are implementations $p \Rightarrow \mathsf{Free}\,q$, composed via Kleisli composition. Let $\mathbf{Spec}$ (the category of \emph{specifications}) be the category whose objects are pairs $(p, r)$ of a polynomial and a dependent polynomial, and whose morphisms $(p, r) \to (q, s)$ are pairs $(f, \hat{f})$ of an implementation and a verification of that implementation.

Both categories carry monoidal structures given by \(\oplus\), and the
forgetful functor \(\pi : \mathbf{Spec} \to \mathbf{Int}\) is (strictly)
monoidal; this is precisely why verified implementations compose
compatibly with wiring diagrams.

The assignment \(p \mapsto \mathsf{Mealy}\,p\) then defines a
\emph{presheaf} on \(\mathbf{Int}\), i.e.~a contravariant functor
\(\mathsf{Mealy} : \mathbf{Int}^{\mathrm{op}} \to \mathbf{Set}\), with
the functorial action given by \texttt{prog→mealy}. Moreover, this
presheaf is \emph{lax monoidal}, since the operations \texttt{mealy⊕}
and \texttt{mealyzero} provide natural maps
\[\mathsf{Mealy}(p) \times \mathsf{Mealy}(q) \to \mathsf{Mealy}(p \oplus q) \qquad \text{and} \qquad 1 \to \mathsf{Mealy}(\mathsf{zeroPoly})\]
satisfying the coherence conditions of a lax monoidal functor.

Similarly, \(\mathsf{DepMealy}\) defines a lax monoidal presheaf on
\(\mathbf{Spec}^{\mathrm{op}}\) that maps a dependent polynomial
\((p,r)\) to the set of Mealy machines with interface \(p\) satisfying
the specification \(r\), and the projection map
\(\mathsf{DepMealy}(p, r) \to \mathsf{Mealy}(p)\) is then a monoidal
natural transformation
\(\mathsf{DepMealy} \Rightarrow \mathsf{Mealy} \circ \pi\).

The most general abstract setting for compositional verification is then
as follows:

\begin{definition}\label{def:abstract-framework}
A \emph{compositional verification framework} consists of: (i) a monoidal category $(\mathbf{Int}, \oplus, 0)$ with a lax monoidal presheaf $\mathsf{Sys} : \mathbf{Int}^{\mathrm{op}} \to \mathbf{Set}$ of \emph{systems}; (ii) a monoidal category $(\mathbf{Spec}, \oplus, 0)$ with a monoidal functor $\pi : \mathbf{Spec} \to \mathbf{Int}$ and a lax monoidal presheaf $\mathsf{Verif} : \mathbf{Spec}^{\mathrm{op}} \to \mathbf{Set}$ of \emph{verified systems}; (iii) a monoidal natural transformation $\mathsf{Verif} \Rightarrow \mathsf{Sys} \circ \pi$.
\end{definition}

Our concrete development instantiates this with \(\mathbf{Int}\) as the
Kleisli category of the free monad on \(\mathsf{Poly}\),
\(\mathbf{Spec}\) as the corresponding category of dependent
polynomials, \(\mathsf{Sys} = \mathsf{Mealy}\), and
\(\mathsf{Verif} = \mathsf{DepMealy}\). The value of the abstraction is
that it identifies precisely what structure is needed for compositional
verification, and opens the door to other instantiations. In particular,
the sequential modules considered so far can be extended to
\emph{concurrent} modules by replacing the coproduct \(\oplus\) with a
\emph{parallel sum} \(p \| q\) that allows calling either or both
interfaces simultaneously. This yields a second monoidal structure on
\(\mathbf{Int}\). Details are given in Appendix A.

\section{Related Work}\label{related-work}

The theory of polynomial functors as a framework for interaction has
been developed by Spivak and collaborators (Niu and Spivak 2024; Spivak
2022; Spivak and Tan 2017), and Libkind and Spivak (Libkind and Spivak
2025) study the free monad--cofree comonad duality (see Appendix B). The
connection of polynomial functors to dependent type theory has been
explored through \emph{containers} (Abbott et al. 2005) and W-types
(Gambino and Hyland 2004); our dependent polynomials extend this with a
specification layer.

\section{Conclusion and Future Work}\label{conclusion-and-future-work}

We have presented a framework for compositional program verification
based on polynomial functors in dependent type theory, providing a
uniform treatment of interfaces, implementations, specifications, and
operational semantics, all composing along wiring diagrams, and unified
under an account of the abstract categorical structure behind such
compositionality.

Several directions for future work suggest themselves. First, the
compositional nature of the framework makes it highly amenable to
\emph{automation}: the composition operations could be implemented as
tactics or elaboration procedures in a proof assistant, reducing the
burden on the user to supply only the local verifications for individual
modules. Second, while our formalization is in Agda, the general
principles should be applicable in other dependently typed languages
such as Lean, Idris, or Coq, and we are interested in exploring such
implementations. Third, wiring diagrams make \emph{information flow}
explicit in a program architecture, and we are interested in applying
this to the verification of security properties and information flow
policies. In a world where digital systems are increasingly built from
opaque components---API calls, machine learning models---techniques for
formally guaranteeing their safety through compositional,
interface-level reasoning become ever more valuable.

\subsection{Acknowledgements}\label{acknowledgements}

I am grateful to my colleagues at the Topos Institute during my Summers
there as a Research Assistant for many stimulating discussions that
ultimately led to the development of this framework. In particular,
conversations with David Spivak, David Jaz Myers, Evan Patterson, and
Kevin Carlson were instrumental in shaping the ideas presented here, and
my conversations with Dana Scott about Agda likewise inspired the
formalization presented in this paper.

\section*{References}\label{references}
\addcontentsline{toc}{section}{References}

\protect\phantomsection\label{refs}

\appendix

\section{Concurrency and Relational
Verification}\label{concurrency-and-relational-verification}

So far, the program modules we have considered have been
\emph{sequential}, in that they may only call one of their dependencies
at a time. However, our operational semantics in terms of Mealy machines
does not require this, and we can extend our framework to allow for
\emph{concurrent} program modules that may call multiple dependencies at
once.

To this end, we make use of the \emph{conjunctive sum} of types, which
is defined as \(A \vee B = A + B + (A \times B)\), representing the
possibility of either an \(A\), a \(B\), or both at the same time. In
Agda, we can define this as follows:

\begin{Shaded}
\begin{Highlighting}[]
\KeywordTok{data} \OtherTok{\_}\NormalTok{∨}\OtherTok{\_} \OtherTok{(}\NormalTok{A B }\OtherTok{:} \DataTypeTok{Set}\OtherTok{)} \OtherTok{:} \DataTypeTok{Set} \KeywordTok{where}
\NormalTok{    left }\OtherTok{:}\NormalTok{ A }\OtherTok{→}\NormalTok{ A ∨ B}
\NormalTok{    right }\OtherTok{:}\NormalTok{ B }\OtherTok{→}\NormalTok{ A ∨ B}
\NormalTok{    both }\OtherTok{:}\NormalTok{ A }\OtherTok{→}\NormalTok{ B }\OtherTok{→}\NormalTok{ A ∨ B}
\end{Highlighting}
\end{Shaded}

Using this, we define the \emph{parallel sum} of polynomials as follows:

\begin{Shaded}
\begin{Highlighting}[]
\OtherTok{\_}\NormalTok{∣∣}\OtherTok{\_} \OtherTok{:}\NormalTok{ Poly }\OtherTok{→}\NormalTok{ Poly }\OtherTok{→}\NormalTok{ Poly}
\OtherTok{\_}\NormalTok{∣∣}\OtherTok{\_} \OtherTok{(}\NormalTok{A , B}\OtherTok{)} \OtherTok{(}\NormalTok{C , D}\OtherTok{)} \OtherTok{.}\NormalTok{fst }\OtherTok{=}\NormalTok{ A ∨ C}
\OtherTok{\_}\NormalTok{∣∣}\OtherTok{\_} \OtherTok{(}\NormalTok{A , B}\OtherTok{)} \OtherTok{(}\NormalTok{C , D}\OtherTok{)} \OtherTok{.}\NormalTok{snd }\OtherTok{(}\NormalTok{left a}\OtherTok{)} \OtherTok{=}\NormalTok{ B a}
\OtherTok{\_}\NormalTok{∣∣}\OtherTok{\_} \OtherTok{(}\NormalTok{A , B}\OtherTok{)} \OtherTok{(}\NormalTok{C , D}\OtherTok{)} \OtherTok{.}\NormalTok{snd }\OtherTok{(}\NormalTok{right c}\OtherTok{)} \OtherTok{=}\NormalTok{ D c}
\OtherTok{\_}\NormalTok{∣∣}\OtherTok{\_} \OtherTok{(}\NormalTok{A , B}\OtherTok{)} \OtherTok{(}\NormalTok{C , D}\OtherTok{)} \OtherTok{.}\NormalTok{snd }\OtherTok{(}\NormalTok{both a c}\OtherTok{)} \OtherTok{=}\NormalTok{ B a × D c}
\end{Highlighting}
\end{Shaded}

Thinking of this as an interface to a pair of functions/program modules,
what this allows is for us to call either one or both of the
functions/modules at once, and to obtain a corresponding result of the
appropriate type.

The parallel sum of dependent polynomials is defined similarly:

\begin{Shaded}
\begin{Highlighting}[]
\OtherTok{\_}\NormalTok{∣∣Dep}\OtherTok{\_} \OtherTok{:} \OtherTok{\{}\NormalTok{p q }\OtherTok{:}\NormalTok{ Poly}\OtherTok{\}} \OtherTok{→}\NormalTok{ DepPoly p }\OtherTok{→}\NormalTok{ DepPoly q }\OtherTok{→}\NormalTok{ DepPoly }\OtherTok{(}\NormalTok{p ∣∣ q}\OtherTok{)}
\OtherTok{\_}\NormalTok{∣∣Dep}\OtherTok{\_} \OtherTok{(}\NormalTok{A , B}\OtherTok{)} \OtherTok{(}\NormalTok{C , D}\OtherTok{)} \OtherTok{.}\NormalTok{fst }\OtherTok{(}\NormalTok{left a}\OtherTok{)} \OtherTok{=}\NormalTok{ A a}
\OtherTok{\_}\NormalTok{∣∣Dep}\OtherTok{\_} \OtherTok{(}\NormalTok{A , B}\OtherTok{)} \OtherTok{(}\NormalTok{C , D}\OtherTok{)} \OtherTok{.}\NormalTok{fst }\OtherTok{(}\NormalTok{right c}\OtherTok{)} \OtherTok{=}\NormalTok{ C c}
\OtherTok{\_}\NormalTok{∣∣Dep}\OtherTok{\_} \OtherTok{(}\NormalTok{A , B}\OtherTok{)} \OtherTok{(}\NormalTok{C , D}\OtherTok{)} \OtherTok{.}\NormalTok{fst }\OtherTok{(}\NormalTok{both a c}\OtherTok{)} \OtherTok{=}\NormalTok{ A a × C c}
\OtherTok{\_}\NormalTok{∣∣Dep}\OtherTok{\_} \OtherTok{(}\NormalTok{A , B}\OtherTok{)} \OtherTok{(}\NormalTok{C , D}\OtherTok{)} \OtherTok{.}\NormalTok{snd }\OtherTok{(}\NormalTok{left a}\OtherTok{)}\NormalTok{ y }\OtherTok{=}\NormalTok{ B a y}
\OtherTok{\_}\NormalTok{∣∣Dep}\OtherTok{\_} \OtherTok{(}\NormalTok{A , B}\OtherTok{)} \OtherTok{(}\NormalTok{C , D}\OtherTok{)} \OtherTok{.}\NormalTok{snd }\OtherTok{(}\NormalTok{right c}\OtherTok{)}\NormalTok{ y }\OtherTok{=}\NormalTok{ D c y}
\OtherTok{\_}\NormalTok{∣∣Dep}\OtherTok{\_} \OtherTok{(}\NormalTok{A , B}\OtherTok{)} \OtherTok{(}\NormalTok{C , D}\OtherTok{)} \OtherTok{.}\NormalTok{snd }\OtherTok{(}\NormalTok{both a c}\OtherTok{)} \OtherTok{(}\NormalTok{y , z}\OtherTok{)} \OtherTok{(}\NormalTok{x , w}\OtherTok{)} \OtherTok{=} 
    \OtherTok{(}\NormalTok{B a y x}\OtherTok{)}\NormalTok{ × }\OtherTok{(}\NormalTok{D c z w}\OtherTok{)}
\end{Highlighting}
\end{Shaded}

The parallel sum is in fact a (symmetric) monoidal product on the
category of polynomials, with unit \texttt{zeroPoly}. We can exhibit
this by defining the following operations that allow us to put program
modules \emph{in parallel} using the parallel sum:

\begin{Shaded}
\begin{Highlighting}[]
\CommentTok{{-}{-} injections into the parallel sum}
\NormalTok{leftProg }\OtherTok{:} \OtherTok{\{}\NormalTok{p q }\OtherTok{:}\NormalTok{ Poly}\OtherTok{\}} \OtherTok{\{}\NormalTok{E }\OtherTok{:} \DataTypeTok{Set}\OtherTok{\}} \OtherTok{→}\NormalTok{ Free p E }\OtherTok{→}\NormalTok{ Free }\OtherTok{(}\NormalTok{p ∣∣ q}\OtherTok{)}\NormalTok{ E}
\NormalTok{leftProg }\OtherTok{(}\NormalTok{return e}\OtherTok{)} \OtherTok{=}\NormalTok{ return e}
\NormalTok{leftProg }\OtherTok{(}\NormalTok{bind a k}\OtherTok{)} \OtherTok{=}\NormalTok{ bind }\OtherTok{(}\NormalTok{left a}\OtherTok{)} \OtherTok{(λ}\NormalTok{ b }\OtherTok{→}\NormalTok{ leftProg }\OtherTok{(}\NormalTok{k b}\OtherTok{))}

\NormalTok{rightProg }\OtherTok{:} \OtherTok{\{}\NormalTok{p q }\OtherTok{:}\NormalTok{ Poly}\OtherTok{\}} \OtherTok{\{}\NormalTok{F }\OtherTok{:} \DataTypeTok{Set}\OtherTok{\}} \OtherTok{→}\NormalTok{ Free q F }\OtherTok{→}\NormalTok{ Free }\OtherTok{(}\NormalTok{p ∣∣ q}\OtherTok{)}\NormalTok{ F}
\NormalTok{rightProg }\OtherTok{(}\NormalTok{return f}\OtherTok{)} \OtherTok{=}\NormalTok{ return f}
\NormalTok{rightProg }\OtherTok{(}\NormalTok{bind c h}\OtherTok{)} \OtherTok{=}\NormalTok{ bind }\OtherTok{(}\NormalTok{right c}\OtherTok{)} \OtherTok{(λ}\NormalTok{ d }\OtherTok{→}\NormalTok{ rightProg }\OtherTok{(}\NormalTok{h d}\OtherTok{))}

\CommentTok{{-}{-} parallel composition of programs}
\NormalTok{bothProg }\OtherTok{:} \OtherTok{\{}\NormalTok{p q }\OtherTok{:}\NormalTok{ Poly}\OtherTok{\}} \OtherTok{\{}\NormalTok{E F }\OtherTok{:} \DataTypeTok{Set}\OtherTok{\}}
         \OtherTok{→}\NormalTok{ Free p E }\OtherTok{→}\NormalTok{ Free q F }\OtherTok{→}\NormalTok{ Free }\OtherTok{(}\NormalTok{p ∣∣ q}\OtherTok{)} \OtherTok{(}\NormalTok{E × F}\OtherTok{)}
\NormalTok{bothProg }\OtherTok{(}\NormalTok{return e}\OtherTok{)} \OtherTok{(}\NormalTok{return f}\OtherTok{)} \OtherTok{=}\NormalTok{ return }\OtherTok{(}\NormalTok{e , f}\OtherTok{)}
\NormalTok{bothProg }\OtherTok{(}\NormalTok{bind a k}\OtherTok{)} \OtherTok{(}\NormalTok{return f}\OtherTok{)} \OtherTok{=} 
\NormalTok{    bind }\OtherTok{(}\NormalTok{left a}\OtherTok{)} \OtherTok{(λ}\NormalTok{ b }\OtherTok{→}\NormalTok{ bothProg }\OtherTok{(}\NormalTok{k b}\OtherTok{)} \OtherTok{(}\NormalTok{return f}\OtherTok{))}
\NormalTok{bothProg }\OtherTok{(}\NormalTok{return e}\OtherTok{)} \OtherTok{(}\NormalTok{bind c h}\OtherTok{)} \OtherTok{=} 
\NormalTok{    bind }\OtherTok{(}\NormalTok{right c}\OtherTok{)} \OtherTok{(λ}\NormalTok{ d }\OtherTok{→}\NormalTok{ bothProg }\OtherTok{(}\NormalTok{return e}\OtherTok{)} \OtherTok{(}\NormalTok{h d}\OtherTok{))}
\NormalTok{bothProg }\OtherTok{(}\NormalTok{bind a k}\OtherTok{)} \OtherTok{(}\NormalTok{bind c h}\OtherTok{)} \OtherTok{=} 
\NormalTok{    bind }\OtherTok{(}\NormalTok{both a c}\OtherTok{)} \OtherTok{(λ} \OtherTok{(}\NormalTok{b , d}\OtherTok{)} \OtherTok{→}\NormalTok{ bothProg }\OtherTok{(}\NormalTok{k b}\OtherTok{)} \OtherTok{(}\NormalTok{h d}\OtherTok{))}

\CommentTok{{-}{-} parallel composition of program modules}
\OtherTok{\_}\NormalTok{∣∣Prog}\OtherTok{\_} \OtherTok{:} \OtherTok{\{}\NormalTok{p q r s }\OtherTok{:}\NormalTok{ Poly}\OtherTok{\}} \OtherTok{→} \OtherTok{(}\NormalTok{p ⇒ Free r}\OtherTok{)} \OtherTok{→} \OtherTok{(}\NormalTok{q ⇒ Free s}\OtherTok{)} 
         \OtherTok{→} \OtherTok{((}\NormalTok{p ∣∣ q}\OtherTok{)}\NormalTok{ ⇒ Free }\OtherTok{(}\NormalTok{r ∣∣ s}\OtherTok{))}
\OtherTok{\_}\NormalTok{∣∣Prog}\OtherTok{\_}\NormalTok{ f g }\OtherTok{(}\NormalTok{left a}\OtherTok{)} \OtherTok{=}\NormalTok{ leftProg }\OtherTok{(}\NormalTok{f a}\OtherTok{)}
\OtherTok{\_}\NormalTok{∣∣Prog}\OtherTok{\_}\NormalTok{ f g }\OtherTok{(}\NormalTok{right c}\OtherTok{)} \OtherTok{=}\NormalTok{ rightProg }\OtherTok{(}\NormalTok{g c}\OtherTok{)}
\OtherTok{\_}\NormalTok{∣∣Prog}\OtherTok{\_}\NormalTok{ f g }\OtherTok{(}\NormalTok{both a c}\OtherTok{)} \OtherTok{=}\NormalTok{ bothProg }\OtherTok{(}\NormalTok{f a}\OtherTok{)} \OtherTok{(}\NormalTok{g c}\OtherTok{)}
\end{Highlighting}
\end{Shaded}

And similarly for dependent polynomial morphisms/verifications:

\begin{Shaded}
\begin{Highlighting}[]
\CommentTok{{-}{-} injections into the parallel sum of dependent polynomials}
\NormalTok{leftDep }\OtherTok{:} \OtherTok{\{}\NormalTok{p q }\OtherTok{:}\NormalTok{ Poly}\OtherTok{\}} \OtherTok{\{}\NormalTok{r }\OtherTok{:}\NormalTok{ DepPoly p}\OtherTok{\}} \OtherTok{\{}\NormalTok{s }\OtherTok{:}\NormalTok{ DepPoly q}\OtherTok{\}} 
        \OtherTok{→} \OtherTok{\{}\NormalTok{X }\OtherTok{:} \DataTypeTok{Set}\OtherTok{\}} \OtherTok{\{}\NormalTok{Y }\OtherTok{:}\NormalTok{ X }\OtherTok{→} \DataTypeTok{Set}\OtherTok{\}}
        \OtherTok{→} \OtherTok{\{}\NormalTok{f }\OtherTok{:}\NormalTok{ Free p X}\OtherTok{\}}
        \OtherTok{→}\NormalTok{ FreeDep p r X Y f }
        \OtherTok{→}\NormalTok{ FreeDep }\OtherTok{(}\NormalTok{p ∣∣ q}\OtherTok{)} \OtherTok{(}\NormalTok{r ∣∣Dep s}\OtherTok{)}\NormalTok{ X Y }\OtherTok{(}\NormalTok{leftProg f}\OtherTok{)}
\NormalTok{leftDep }\OtherTok{(}\NormalTok{returnD f}\OtherTok{)} \OtherTok{=}\NormalTok{ returnD f}
\NormalTok{leftDep }\OtherTok{(}\NormalTok{bindD c h}\OtherTok{)} \OtherTok{=}\NormalTok{ bindD c }\OtherTok{(λ}\NormalTok{ d e }\OtherTok{→}\NormalTok{ leftDep }\OtherTok{(}\NormalTok{h d e}\OtherTok{))}

\NormalTok{rightDep }\OtherTok{:} \OtherTok{\{}\NormalTok{p q }\OtherTok{:}\NormalTok{ Poly}\OtherTok{\}} \OtherTok{\{}\NormalTok{r }\OtherTok{:}\NormalTok{ DepPoly p}\OtherTok{\}} \OtherTok{\{}\NormalTok{s }\OtherTok{:}\NormalTok{ DepPoly q}\OtherTok{\}} 
         \OtherTok{→} \OtherTok{\{}\NormalTok{X }\OtherTok{:} \DataTypeTok{Set}\OtherTok{\}} \OtherTok{\{}\NormalTok{Y }\OtherTok{:}\NormalTok{ X }\OtherTok{→} \DataTypeTok{Set}\OtherTok{\}}
         \OtherTok{→} \OtherTok{\{}\NormalTok{g }\OtherTok{:}\NormalTok{ Free q X}\OtherTok{\}}
         \OtherTok{→}\NormalTok{ FreeDep q s X Y g }
         \OtherTok{→}\NormalTok{ FreeDep }\OtherTok{(}\NormalTok{p ∣∣ q}\OtherTok{)} \OtherTok{(}\NormalTok{r ∣∣Dep s}\OtherTok{)}\NormalTok{ X Y }\OtherTok{(}\NormalTok{rightProg g}\OtherTok{)}
\NormalTok{rightDep }\OtherTok{(}\NormalTok{returnD f}\OtherTok{)} \OtherTok{=}\NormalTok{ returnD f}
\NormalTok{rightDep }\OtherTok{(}\NormalTok{bindD c h}\OtherTok{)} \OtherTok{=}\NormalTok{ bindD c }\OtherTok{(λ}\NormalTok{ d e }\OtherTok{→}\NormalTok{ rightDep }\OtherTok{(}\NormalTok{h d e}\OtherTok{))}

\CommentTok{{-}{-} parallel composition of verifications}
\NormalTok{bothDep }\OtherTok{:} \OtherTok{\{}\NormalTok{p q }\OtherTok{:}\NormalTok{ Poly}\OtherTok{\}} \OtherTok{\{}\NormalTok{r }\OtherTok{:}\NormalTok{ DepPoly p}\OtherTok{\}} \OtherTok{\{}\NormalTok{s }\OtherTok{:}\NormalTok{ DepPoly q}\OtherTok{\}} 
        \OtherTok{→} \OtherTok{\{}\NormalTok{X Y }\OtherTok{:} \DataTypeTok{Set}\OtherTok{\}} \OtherTok{\{}\NormalTok{Z }\OtherTok{:}\NormalTok{ X }\OtherTok{→} \DataTypeTok{Set}\OtherTok{\}} \OtherTok{\{}\NormalTok{W }\OtherTok{:}\NormalTok{ Y }\OtherTok{→} \DataTypeTok{Set}\OtherTok{\}}
        \OtherTok{→} \OtherTok{\{}\NormalTok{f }\OtherTok{:}\NormalTok{ Free p X}\OtherTok{\}} \OtherTok{\{}\NormalTok{g }\OtherTok{:}\NormalTok{ Free q Y}\OtherTok{\}}
        \OtherTok{→}\NormalTok{ FreeDep p r X Z f }\OtherTok{→}\NormalTok{ FreeDep q s Y W g }
        \OtherTok{→}\NormalTok{ FreeDep }\OtherTok{(}\NormalTok{p ∣∣ q}\OtherTok{)} \OtherTok{(}\NormalTok{r ∣∣Dep s}\OtherTok{)} \OtherTok{(}\NormalTok{X × Y}\OtherTok{)} 
                  \OtherTok{(λ} \OtherTok{(}\NormalTok{x , y}\OtherTok{)} \OtherTok{→}\NormalTok{ Z x × W y}\OtherTok{)} \OtherTok{(}\NormalTok{bothProg f g}\OtherTok{)}
\NormalTok{bothDep }\OtherTok{(}\NormalTok{returnD z}\OtherTok{)} \OtherTok{(}\NormalTok{returnD w}\OtherTok{)} \OtherTok{=}\NormalTok{ returnD }\OtherTok{(}\NormalTok{z , w}\OtherTok{)}
\NormalTok{bothDep }\OtherTok{(}\NormalTok{bindD c h}\OtherTok{)} \OtherTok{(}\NormalTok{returnD w}\OtherTok{)} \OtherTok{=} 
\NormalTok{    bindD c }\OtherTok{(λ}\NormalTok{ b x }\OtherTok{→}\NormalTok{ bothDep }\OtherTok{(}\NormalTok{h b x}\OtherTok{)} \OtherTok{(}\NormalTok{returnD w}\OtherTok{))}
\NormalTok{bothDep }\OtherTok{(}\NormalTok{returnD z}\OtherTok{)} \OtherTok{(}\NormalTok{bindD d k}\OtherTok{)} \OtherTok{=} 
\NormalTok{    bindD d }\OtherTok{(λ}\NormalTok{ b x }\OtherTok{→}\NormalTok{ bothDep }\OtherTok{(}\NormalTok{returnD z}\OtherTok{)} \OtherTok{(}\NormalTok{k b x}\OtherTok{))}
\NormalTok{bothDep }\OtherTok{(}\NormalTok{bindD c h}\OtherTok{)} \OtherTok{(}\NormalTok{bindD d k}\OtherTok{)} \OtherTok{=} 
\NormalTok{    bindD }\OtherTok{(}\NormalTok{c , d}\OtherTok{)} \OtherTok{(λ} \OtherTok{(}\NormalTok{a , b}\OtherTok{)} \OtherTok{(}\NormalTok{x , y}\OtherTok{)} \OtherTok{→}\NormalTok{ bothDep }\OtherTok{(}\NormalTok{h a x}\OtherTok{)} \OtherTok{(}\NormalTok{k b y}\OtherTok{))}

\CommentTok{{-}{-} parallel composition of dependent polynomial morphisms}
\OtherTok{\_}\NormalTok{∣∣DepProg}\OtherTok{\_} \OtherTok{:} \OtherTok{\{}\NormalTok{p q r s }\OtherTok{:}\NormalTok{ Poly}\OtherTok{\}} 
            \OtherTok{→} \OtherTok{\{}\NormalTok{p\textquotesingle{} }\OtherTok{:}\NormalTok{ DepPoly p}\OtherTok{\}} \OtherTok{\{}\NormalTok{q\textquotesingle{} }\OtherTok{:}\NormalTok{ DepPoly q}\OtherTok{\}} 
            \OtherTok{→} \OtherTok{\{}\NormalTok{r\textquotesingle{} }\OtherTok{:}\NormalTok{ DepPoly r}\OtherTok{\}} \OtherTok{\{}\NormalTok{s\textquotesingle{} }\OtherTok{:}\NormalTok{ DepPoly s}\OtherTok{\}}
            \OtherTok{→} \OtherTok{\{}\NormalTok{f }\OtherTok{:}\NormalTok{ p ⇒ Free r}\OtherTok{\}} \OtherTok{\{}\NormalTok{g }\OtherTok{:}\NormalTok{ q ⇒ Free s}\OtherTok{\}}
            \OtherTok{→}\NormalTok{ ⇒Dep p p\textquotesingle{} }\OtherTok{(}\NormalTok{FreeDep r r\textquotesingle{}}\OtherTok{)}\NormalTok{ f }\OtherTok{→}\NormalTok{ ⇒Dep q q\textquotesingle{} }\OtherTok{(}\NormalTok{FreeDep s s\textquotesingle{}}\OtherTok{)}\NormalTok{ g}
            \OtherTok{→}\NormalTok{ ⇒Dep }\OtherTok{(}\NormalTok{p ∣∣ q}\OtherTok{)} \OtherTok{(}\NormalTok{p\textquotesingle{} ∣∣Dep q\textquotesingle{}}\OtherTok{)} 
                   \OtherTok{(}\NormalTok{FreeDep }\OtherTok{(}\NormalTok{r ∣∣ s}\OtherTok{)} \OtherTok{(}\NormalTok{r\textquotesingle{} ∣∣Dep s\textquotesingle{}}\OtherTok{))} \OtherTok{(}\NormalTok{f ∣∣Prog g}\OtherTok{)}
\OtherTok{\_}\NormalTok{∣∣DepProg}\OtherTok{\_}\NormalTok{ ff gg }\OtherTok{(}\NormalTok{left a}\OtherTok{)}\NormalTok{ b }\OtherTok{=}\NormalTok{ leftDep }\OtherTok{(}\NormalTok{ff a b}\OtherTok{)}
\OtherTok{\_}\NormalTok{∣∣DepProg}\OtherTok{\_}\NormalTok{ ff gg }\OtherTok{(}\NormalTok{right c}\OtherTok{)}\NormalTok{ d }\OtherTok{=}\NormalTok{ rightDep }\OtherTok{(}\NormalTok{gg c d}\OtherTok{)}
\OtherTok{\_}\NormalTok{∣∣DepProg}\OtherTok{\_}\NormalTok{ ff gg }\OtherTok{(}\NormalTok{both a c}\OtherTok{)} \OtherTok{(}\NormalTok{b , d}\OtherTok{)} \OtherTok{=}\NormalTok{ bothDep }\OtherTok{(}\NormalTok{ff a b}\OtherTok{)} \OtherTok{(}\NormalTok{gg c d}\OtherTok{)}
\end{Highlighting}
\end{Shaded}

Hence, since our generalized account of compositional verification is
based on the existence of a monoidal structure on the category of
interfaces and the category of specifications, we can apply this same
framework for compositional verification of concurrent program modules,
by simply using the parallel sum as the monoidal structure instead of
the coproduct.

To complete the picture, we also show that Mealy machines and their
verifications are closed under the parallel sum, allowing for Mealy
machines to genuinely be run in parallel:

\begin{Shaded}
\begin{Highlighting}[]
\CommentTok{{-}{-} parallel sum of Mealy machines}
\OtherTok{\_}\NormalTok{mealy||}\OtherTok{\_} \OtherTok{:} \OtherTok{\{}\NormalTok{p q }\OtherTok{:}\NormalTok{ Poly}\OtherTok{\}} \OtherTok{→}\NormalTok{ Mealy p }\OtherTok{→}\NormalTok{ Mealy q }\OtherTok{→}\NormalTok{ Mealy }\OtherTok{(}\NormalTok{p ∣∣ q}\OtherTok{)}
\OtherTok{\_}\NormalTok{mealy||}\OtherTok{\_}\NormalTok{ m₁ m₂ }\OtherTok{.}\NormalTok{app }\OtherTok{(}\NormalTok{left a}\OtherTok{)} \OtherTok{=} 
    \KeywordTok{let} \OtherTok{(}\NormalTok{b , m\textquotesingle{}}\OtherTok{)} \OtherTok{=}\NormalTok{ m₁ }\OtherTok{.}\NormalTok{app a }\KeywordTok{in} 
    \OtherTok{(}\NormalTok{b , m\textquotesingle{} mealy|| m₂}\OtherTok{)}
\OtherTok{\_}\NormalTok{mealy||}\OtherTok{\_}\NormalTok{ m₁ m₂ }\OtherTok{.}\NormalTok{app }\OtherTok{(}\NormalTok{right c}\OtherTok{)} \OtherTok{=} 
    \KeywordTok{let} \OtherTok{(}\NormalTok{d , m\textquotesingle{}}\OtherTok{)} \OtherTok{=}\NormalTok{ m₂ }\OtherTok{.}\NormalTok{app c }\KeywordTok{in} 
    \OtherTok{(}\NormalTok{d , m₁ mealy|| m\textquotesingle{}}\OtherTok{)}
\OtherTok{\_}\NormalTok{mealy||}\OtherTok{\_}\NormalTok{ m₁ m₂ }\OtherTok{.}\NormalTok{app }\OtherTok{(}\NormalTok{both a c}\OtherTok{)} \OtherTok{=} 
    \KeywordTok{let} \OtherTok{((}\NormalTok{b , m₁\textquotesingle{}}\OtherTok{)}\NormalTok{ , }\OtherTok{(}\NormalTok{d , m₂\textquotesingle{}}\OtherTok{))} \OtherTok{=} \OtherTok{(}\NormalTok{m₁ }\OtherTok{.}\NormalTok{app a , m₂ }\OtherTok{.}\NormalTok{app c}\OtherTok{)} \KeywordTok{in} 
    \OtherTok{((}\NormalTok{b , d}\OtherTok{)}\NormalTok{ , m₁\textquotesingle{} mealy|| m₂\textquotesingle{}}\OtherTok{)}

\CommentTok{{-}{-} parallel sum of dependent Mealy machines}
\OtherTok{\_}\NormalTok{depMealy||}\OtherTok{\_} \OtherTok{:} \OtherTok{\{}\NormalTok{p q }\OtherTok{:}\NormalTok{ Poly}\OtherTok{\}} \OtherTok{\{}\NormalTok{r }\OtherTok{:}\NormalTok{ DepPoly p}\OtherTok{\}} \OtherTok{\{}\NormalTok{s }\OtherTok{:}\NormalTok{ DepPoly q}\OtherTok{\}}
             \OtherTok{→} \OtherTok{\{}\NormalTok{m₁ }\OtherTok{:}\NormalTok{ Mealy p}\OtherTok{\}} \OtherTok{\{}\NormalTok{m₂ }\OtherTok{:}\NormalTok{ Mealy q}\OtherTok{\}}
             \OtherTok{→}\NormalTok{ DepMealy p r m₁ }\OtherTok{→}\NormalTok{ DepMealy q s m₂ }
             \OtherTok{→}\NormalTok{ DepMealy }\OtherTok{(}\NormalTok{p ∣∣ q}\OtherTok{)} \OtherTok{(}\NormalTok{r ∣∣Dep s}\OtherTok{)} \OtherTok{(}\NormalTok{m₁ mealy|| m₂}\OtherTok{)}
\OtherTok{\_}\NormalTok{depMealy||}\OtherTok{\_}\NormalTok{ m₁ m₂ }\OtherTok{.}\NormalTok{appD }\OtherTok{(}\NormalTok{left a}\OtherTok{)}\NormalTok{ c }\OtherTok{=} 
    \KeywordTok{let} \OtherTok{(}\NormalTok{b , m\textquotesingle{}}\OtherTok{)} \OtherTok{=}\NormalTok{ m₁ }\OtherTok{.}\NormalTok{appD a c }\KeywordTok{in} 
    \OtherTok{(}\NormalTok{b , }\OtherTok{(}\NormalTok{m\textquotesingle{} depMealy|| m₂}\OtherTok{))}
\OtherTok{\_}\NormalTok{depMealy||}\OtherTok{\_}\NormalTok{ m₁ m₂ }\OtherTok{.}\NormalTok{appD }\OtherTok{(}\NormalTok{right c}\OtherTok{)}\NormalTok{ d }\OtherTok{=} 
    \KeywordTok{let} \OtherTok{(}\NormalTok{b , m\textquotesingle{}}\OtherTok{)} \OtherTok{=}\NormalTok{ m₂ }\OtherTok{.}\NormalTok{appD c d }\KeywordTok{in} 
    \OtherTok{(}\NormalTok{b , }\OtherTok{(}\NormalTok{m₁ depMealy|| m\textquotesingle{}}\OtherTok{))}
\OtherTok{\_}\NormalTok{depMealy||}\OtherTok{\_}\NormalTok{ m₁ m₂ }\OtherTok{.}\NormalTok{appD }\OtherTok{(}\NormalTok{both a c}\OtherTok{)} \OtherTok{(}\NormalTok{b , d}\OtherTok{)} \OtherTok{=} 
    \KeywordTok{let} \OtherTok{((}\NormalTok{e , m₁\textquotesingle{}}\OtherTok{)}\NormalTok{ , }\OtherTok{(}\NormalTok{f , m₂\textquotesingle{}}\OtherTok{))} \OtherTok{=} \OtherTok{(}\NormalTok{m₁ }\OtherTok{.}\NormalTok{appD a b , m₂ }\OtherTok{.}\NormalTok{appD c d}\OtherTok{)} \KeywordTok{in} 
    \OtherTok{((}\NormalTok{e , f}\OtherTok{)}\NormalTok{ , }\OtherTok{(}\NormalTok{m₁\textquotesingle{} depMealy|| m₂\textquotesingle{}}\OtherTok{))}
\end{Highlighting}
\end{Shaded}

Hence we have another instance of our general framework for
compositional verification, this time for concurrent program modules and
their operational semantics via parallel composition of Mealy machines.

\subsection{Affine vs.~Cocartesian Structure and
Race-Freedom}\label{affine-vs.-cocartesian-structure-and-race-freedom}

An important distinction between the coproduct \(\oplus\) and the
parallel sum \(\|\) is that \(\oplus\) is \emph{cocartesian}---it admits
a codiagonal \(p \oplus p \to p\)---while \(\|\) is only \emph{affine
symmetric monoidal} (it admits weakening but not contraction, in the
jargon of linear logic/sequent calculus). Concretely, this means that in
a parallel wiring diagram, a wire cannot be used in two places
simultaneously (though it can remain unused, i.e.~it can be left
dangling). In terms of concurrency, this suffices to prevent \emph{race
conditions} with respect to the effects encoded by polynomial
interfaces: by design, concurrent programs structured via the parallel
sum cannot invoke the same effect in two places simultaneously, and
hence there is no ambiguity or nondeterminism as to whether one such
invocation will resolve before another. In other words, any
effect/interface that is used in such a wiring diagram is only ever
\emph{in one place} at a time.

\subsection{Relational Specifications}\label{relational-specifications}

A key feature of the parallel sum is that it enables \emph{relational
verification}. Observe that
\(p \| q \cong p \oplus q \oplus (p \otimes q)\), where \(\otimes\) is
the \emph{parallel product}:

\begin{Shaded}
\begin{Highlighting}[]
\OtherTok{\_}\NormalTok{⊗}\OtherTok{\_} \OtherTok{:}\NormalTok{ Poly }\OtherTok{→}\NormalTok{ Poly }\OtherTok{→}\NormalTok{ Poly}
\OtherTok{(}\NormalTok{A , B}\OtherTok{)}\NormalTok{ ⊗ }\OtherTok{(}\NormalTok{C , D}\OtherTok{)} \OtherTok{=} \OtherTok{(}\NormalTok{A × C , }\OtherTok{λ} \OtherTok{(}\NormalTok{a , c}\OtherTok{)} \OtherTok{→}\NormalTok{ B a × D c}\OtherTok{)}
\end{Highlighting}
\end{Shaded}

A dependent polynomial over \(p \| q\) therefore decomposes into three
parts: a specification for \(p\) alone, a specification for \(q\) alone,
and a \emph{relational} specification over \(p \otimes q\) whose
precondition may relate the inputs to \(p\) and \(q\), and whose
postcondition may relate their outputs. This enables verification of
properties such as \emph{monotonicity} (if the input to one module
increases, the output of another increases correspondingly),
\emph{noninterference} (the output of one module is independent of the
input to another), and other relational properties that cannot be
expressed as unary pre/postconditions.

\section{Connection to ``Pattern Runs on
Matter''}\label{connection-to-pattern-runs-on-matter}

The construction \texttt{prog→mealy} in the paper is closely related to
the ``pattern runs on matter'' framework of Libkind and Spivak (Libkind
and Spivak 2025), specifically:

$\mathsf{Free}\,p\,C$ corresponds to $\mathfrak{m}_p$, the free monad on polynomial $p$ (a ``pattern'' or decision tree). The type $\mathsf{Mealy}\,q$ corresponds to $\mathfrak{c}_{[q,\, y]}$, the cofree comonad on the internal hom $[q, y]$ in $\mathbf{Poly}$, viewed as a monoidal category with the Dirichlet product $\otimes$, where $y$ is the monoidal unit.

Given $f : p \to \mathfrak{m}_q$, the derivation proceeds as:
$$p \otimes \mathfrak{c}_{[q,y]} \xrightarrow{f \otimes \mathrm{id}} \mathfrak{m}_q \otimes \mathfrak{c}_{[q,y]} \xrightarrow{\Xi_{q,[q,y]}} \mathfrak{m}_{q \otimes [q,y]} \xrightarrow{\mathfrak{m}_{app}} \mathfrak{m}_{y} \to y$$
where $\Xi$ is the module structure map and $app : q \otimes [q,y] \to y$ is the evaluation map. Transposing gives $\mathfrak{c}_{[q,y]} \to [p, y]$, and applying the co-Kleisli lift for $\mathfrak{c}$, we obtain $\mathfrak{c}_{[q,y]} \to \mathfrak{c}_{[p,y]}$, whose forward part is precisely $\texttt{prog\textrightarrow mealy}\,f$.

\section{Extended Examples}\label{extended-examples}

\subsection{Verified Fold and Append}\label{verified-fold-and-append}

The verification of the \texttt{fold} module captures the general
pattern of proofs by \emph{induction}. Given a specification \(D\) on
the accumulator and a postcondition \(E\) relating accumulator, input
list, and result, the verification \texttt{foldInd} proceeds by
structural induction, mirroring the recursive structure of
\texttt{fold}:

\begin{Shaded}
\begin{Highlighting}[]
\NormalTok{FoldSpec }\OtherTok{:} \OtherTok{(}\NormalTok{A B C }\OtherTok{:} \DataTypeTok{Set}\OtherTok{)} \OtherTok{(}\NormalTok{D }\OtherTok{:}\NormalTok{ A }\OtherTok{→} \DataTypeTok{Set}\OtherTok{)}
         \OtherTok{→} \OtherTok{(}\NormalTok{E }\OtherTok{:} \OtherTok{(}\NormalTok{a }\OtherTok{:}\NormalTok{ A}\OtherTok{)} \OtherTok{→}\NormalTok{ D a }\OtherTok{→}\NormalTok{ List B }\OtherTok{→}\NormalTok{ C }\OtherTok{→} \DataTypeTok{Set}\OtherTok{)}
         \OtherTok{→}\NormalTok{ DepPoly }\OtherTok{(}\NormalTok{Fold A B C}\OtherTok{)}
\NormalTok{FoldSpec A B C D E }\OtherTok{.}\NormalTok{fst }\OtherTok{(}\NormalTok{a , }\OtherTok{\_)} \OtherTok{=}\NormalTok{ D a}
\NormalTok{FoldSpec A B C D E }\OtherTok{.}\NormalTok{snd }\OtherTok{(}\NormalTok{a , bs}\OtherTok{)}\NormalTok{ d c }\OtherTok{=}\NormalTok{ E a d bs c}

\NormalTok{BaseSpec }\OtherTok{:} \OtherTok{(}\NormalTok{A B C }\OtherTok{:} \DataTypeTok{Set}\OtherTok{)} \OtherTok{(}\NormalTok{D }\OtherTok{:}\NormalTok{ A }\OtherTok{→} \DataTypeTok{Set}\OtherTok{)}
         \OtherTok{→} \OtherTok{(}\NormalTok{E }\OtherTok{:} \OtherTok{(}\NormalTok{a }\OtherTok{:}\NormalTok{ A}\OtherTok{)} \OtherTok{→}\NormalTok{ D a }\OtherTok{→}\NormalTok{ List B }\OtherTok{→}\NormalTok{ C }\OtherTok{→} \DataTypeTok{Set}\OtherTok{)}
         \OtherTok{→}\NormalTok{ DepPoly }\OtherTok{(}\NormalTok{Base A C}\OtherTok{)}
\NormalTok{BaseSpec A B C D E }\OtherTok{.}\NormalTok{fst a }\OtherTok{=}\NormalTok{ D a}
\NormalTok{BaseSpec A B C D E }\OtherTok{.}\NormalTok{snd a d c }\OtherTok{=}\NormalTok{ E a d nil c}

\NormalTok{StepSpec }\OtherTok{:} \OtherTok{(}\NormalTok{A B C }\OtherTok{:} \DataTypeTok{Set}\OtherTok{)} \OtherTok{(}\NormalTok{D }\OtherTok{:}\NormalTok{ A }\OtherTok{→} \DataTypeTok{Set}\OtherTok{)}
         \OtherTok{→} \OtherTok{(}\NormalTok{E }\OtherTok{:} \OtherTok{(}\NormalTok{a }\OtherTok{:}\NormalTok{ A}\OtherTok{)} \OtherTok{→}\NormalTok{ D a }\OtherTok{→}\NormalTok{ List B }\OtherTok{→}\NormalTok{ C }\OtherTok{→} \DataTypeTok{Set}\OtherTok{)}
         \OtherTok{→}\NormalTok{ DepPoly }\OtherTok{(}\NormalTok{Step A B C}\OtherTok{)}
\NormalTok{StepSpec A B C D E }\OtherTok{.}\NormalTok{fst }\OtherTok{(}\NormalTok{a , }\OtherTok{\_}\NormalTok{ , c}\OtherTok{)} \OtherTok{=}
\NormalTok{    Σ }\OtherTok{(}\NormalTok{D a}\OtherTok{)} \OtherTok{(λ}\NormalTok{ d }\OtherTok{→}\NormalTok{ Σ }\OtherTok{(}\NormalTok{List B}\OtherTok{)} \OtherTok{(λ}\NormalTok{ bs }\OtherTok{→}\NormalTok{ E a d bs c}\OtherTok{))}
\NormalTok{StepSpec A B C D E }\OtherTok{.}\NormalTok{snd }\OtherTok{(}\NormalTok{a , b , }\OtherTok{\_)} \OtherTok{(}\NormalTok{d , bs , }\OtherTok{\_)}\NormalTok{ c }\OtherTok{=}
\NormalTok{    E a d }\OtherTok{(}\NormalTok{cons b bs}\OtherTok{)}\NormalTok{ c}

\NormalTok{foldInd }\OtherTok{:} \OtherTok{\{}\NormalTok{A B C }\OtherTok{:} \DataTypeTok{Set}\OtherTok{\}} \OtherTok{\{}\NormalTok{D }\OtherTok{:}\NormalTok{ A }\OtherTok{→} \DataTypeTok{Set}\OtherTok{\}}
        \OtherTok{→} \OtherTok{\{}\NormalTok{E }\OtherTok{:} \OtherTok{(}\NormalTok{a }\OtherTok{:}\NormalTok{ A}\OtherTok{)} \OtherTok{→}\NormalTok{ D a }\OtherTok{→}\NormalTok{ List B }\OtherTok{→}\NormalTok{ C }\OtherTok{→} \DataTypeTok{Set}\OtherTok{\}}
        \OtherTok{→}\NormalTok{ ⇒Dep }\OtherTok{\_} \OtherTok{(}\NormalTok{FoldSpec A B C D E}\OtherTok{)}
               \OtherTok{(}\NormalTok{FreeDep }\OtherTok{\_} \OtherTok{(}\NormalTok{sumDep foldlabels }\OtherTok{\_}
                        \OtherTok{(λ\{}\NormalTok{ base }\OtherTok{→}\NormalTok{ BaseSpec A B C D E}
                          \OtherTok{;}\NormalTok{ step }\OtherTok{→}\NormalTok{ StepSpec A B C D E}\OtherTok{\})))}
\NormalTok{               fold}
\NormalTok{foldInd }\OtherTok{(\_}\NormalTok{ , nil}\OtherTok{)}\NormalTok{ d }\OtherTok{=}
\NormalTok{    bindD d }\OtherTok{(λ} \OtherTok{\_}\NormalTok{ e }\OtherTok{→}\NormalTok{ returnD e}\OtherTok{)}
\NormalTok{foldInd }\OtherTok{(}\NormalTok{a , cons b bs}\OtherTok{)}\NormalTok{ d }\OtherTok{=}
\NormalTok{    \textgreater{}\textgreater{}=Dep }\OtherTok{\_} \OtherTok{(}\NormalTok{foldInd }\OtherTok{(}\NormalTok{a , bs}\OtherTok{)}\NormalTok{ d}\OtherTok{)} \OtherTok{(λ} \OtherTok{\_}\NormalTok{ e }\OtherTok{→}
\NormalTok{        bindD }\OtherTok{(}\NormalTok{d , }\OtherTok{(}\NormalTok{bs , e}\OtherTok{))} \OtherTok{(λ} \OtherTok{\_}\NormalTok{ e\textquotesingle{} }\OtherTok{→}
\NormalTok{            returnD e\textquotesingle{}}\OtherTok{))}
\end{Highlighting}
\end{Shaded}

For append, we define a relational specification and verify the base and
step:

\begin{Shaded}
\begin{Highlighting}[]
\KeywordTok{data}\NormalTok{ Append }\OtherTok{\{}\NormalTok{A }\OtherTok{:} \DataTypeTok{Set}\OtherTok{\}} \OtherTok{:}\NormalTok{ List A }\OtherTok{→}\NormalTok{ ⊤ }\OtherTok{→}\NormalTok{ List A }\OtherTok{→}\NormalTok{ List A }\OtherTok{→} \DataTypeTok{Set} \KeywordTok{where}
\NormalTok{    append{-}nil }\OtherTok{:} \OtherTok{\{}\NormalTok{xs }\OtherTok{:}\NormalTok{ List A}\OtherTok{\}} \OtherTok{→}\NormalTok{ Append xs }\OtherTok{\_}\NormalTok{ nil xs}
\NormalTok{    append{-}cons }\OtherTok{:} \OtherTok{\{}\NormalTok{xs ys zs }\OtherTok{:}\NormalTok{ List A}\OtherTok{\}} \OtherTok{\{}\NormalTok{x }\OtherTok{:}\NormalTok{ A}\OtherTok{\}}
                \OtherTok{→}\NormalTok{ Append ys }\OtherTok{\_}\NormalTok{ xs zs}
                \OtherTok{→}\NormalTok{ Append ys }\OtherTok{\_} \OtherTok{(}\NormalTok{cons x xs}\OtherTok{)} \OtherTok{(}\NormalTok{cons x zs}\OtherTok{)}

\NormalTok{appendBaseSpec }\OtherTok{:} \OtherTok{\{}\NormalTok{A }\OtherTok{:} \DataTypeTok{Set}\OtherTok{\}}
               \OtherTok{→}\NormalTok{ ⇒Dep }\OtherTok{\_} \OtherTok{(}\NormalTok{BaseSpec }\OtherTok{(}\NormalTok{List A}\OtherTok{)}\NormalTok{ A }\OtherTok{(}\NormalTok{List A}\OtherTok{)} \OtherTok{(λ} \OtherTok{\_} \OtherTok{→}\NormalTok{ ⊤}\OtherTok{)}\NormalTok{ Append}\OtherTok{)}
                      \OtherTok{(}\NormalTok{FreeDep }\OtherTok{\_}\NormalTok{ depPolyzero}\OtherTok{)}\NormalTok{ appendBase}
\NormalTok{appendBaseSpec }\OtherTok{\_} \OtherTok{\_} \OtherTok{=}\NormalTok{ returnD append{-}nil}

\NormalTok{appendStepSpec }\OtherTok{:} \OtherTok{\{}\NormalTok{A }\OtherTok{:} \DataTypeTok{Set}\OtherTok{\}}
               \OtherTok{→}\NormalTok{ ⇒Dep }\OtherTok{\_} \OtherTok{(}\NormalTok{StepSpec }\OtherTok{(}\NormalTok{List A}\OtherTok{)}\NormalTok{ A }\OtherTok{(}\NormalTok{List A}\OtherTok{)} \OtherTok{(λ} \OtherTok{\_} \OtherTok{→}\NormalTok{ ⊤}\OtherTok{)}\NormalTok{ Append}\OtherTok{)}
                      \OtherTok{(}\NormalTok{FreeDep }\OtherTok{\_}\NormalTok{ depPolyzero}\OtherTok{)}\NormalTok{ appendStep}
\NormalTok{appendStepSpec }\OtherTok{(\_}\NormalTok{ , }\OtherTok{\_}\NormalTok{ , }\OtherTok{\_)} \OtherTok{(\_}\NormalTok{ , }\OtherTok{(\_}\NormalTok{ , e}\OtherTok{))} \OtherTok{=}\NormalTok{ returnD }\OtherTok{(}\NormalTok{append{-}cons e}\OtherTok{)}

\NormalTok{appendSpec }\OtherTok{:} \OtherTok{\{}\NormalTok{A }\OtherTok{:} \DataTypeTok{Set}\OtherTok{\}}
           \OtherTok{→}\NormalTok{ ⇒Dep }\OtherTok{\_} \OtherTok{(}\NormalTok{FoldSpec }\OtherTok{(}\NormalTok{List A}\OtherTok{)}\NormalTok{ A }\OtherTok{(}\NormalTok{List A}\OtherTok{)} \OtherTok{(λ} \OtherTok{\_} \OtherTok{→}\NormalTok{ ⊤}\OtherTok{)}\NormalTok{ Append}\OtherTok{)}
                  \OtherTok{(}\NormalTok{FreeDep }\OtherTok{\_}\NormalTok{ depPolyzero}\OtherTok{)}\NormalTok{ append}
\NormalTok{appendSpec }\OtherTok{=}
\NormalTok{    compDep }\OtherTok{\_}\NormalTok{ foldInd}
            \OtherTok{(λ\{}\NormalTok{ base }\OtherTok{→}\NormalTok{ appendBaseSpec}
              \OtherTok{;}\NormalTok{ step }\OtherTok{→}\NormalTok{ appendStepSpec}\OtherTok{\})}
\end{Highlighting}
\end{Shaded}

\subsection{Verified Fibonacci}\label{verified-fibonacci}

The verification that \texttt{fib} computes Fibonacci numbers uses a
relational specification and a state invariant:

\begin{Shaded}
\begin{Highlighting}[]
\KeywordTok{data}\NormalTok{ IsFib }\OtherTok{:}\NormalTok{ Nat }\OtherTok{→}\NormalTok{ Nat }\OtherTok{→} \DataTypeTok{Set} \KeywordTok{where}
\NormalTok{    isFibZero }\OtherTok{:}\NormalTok{ IsFib }\DecValTok{0} \DecValTok{0}
\NormalTok{    isFibOne }\OtherTok{:}\NormalTok{ IsFib }\DecValTok{1} \DecValTok{1}
\NormalTok{    isFibRec }\OtherTok{:} \OtherTok{\{}\NormalTok{x y z }\OtherTok{:}\NormalTok{ Nat}\OtherTok{\}}
             \OtherTok{→}\NormalTok{ IsFib x y }\OtherTok{→}\NormalTok{ IsFib }\OtherTok{(}\NormalTok{suc x}\OtherTok{)}\NormalTok{ z}
             \OtherTok{→}\NormalTok{ IsFib }\OtherTok{(}\NormalTok{suc }\OtherTok{(}\NormalTok{suc x}\OtherTok{))} \OtherTok{(}\NormalTok{plus y z}\OtherTok{)}

\NormalTok{FibDep }\OtherTok{:}\NormalTok{ DepPoly }\OtherTok{(}\NormalTok{⊤ , }\OtherTok{λ} \OtherTok{\_} \OtherTok{→}\NormalTok{ Nat}\OtherTok{)}
\NormalTok{FibDep }\OtherTok{.}\NormalTok{fst }\OtherTok{\_} \OtherTok{=}\NormalTok{ ⊤}
\NormalTok{FibDep }\OtherTok{.}\NormalTok{snd }\OtherTok{\_} \OtherTok{\_}\NormalTok{ y }\OtherTok{=}\NormalTok{ Σ Nat }\OtherTok{(λ}\NormalTok{ x }\OtherTok{→}\NormalTok{ IsFib x y}\OtherTok{)}

\NormalTok{FibInvariant }\OtherTok{:} \OtherTok{(}\NormalTok{Nat × Nat}\OtherTok{)} \OtherTok{→} \DataTypeTok{Set}
\NormalTok{FibInvariant }\OtherTok{(}\NormalTok{y , z}\OtherTok{)} \OtherTok{=}
\NormalTok{    Σ Nat }\OtherTok{(λ}\NormalTok{ x }\OtherTok{→}\NormalTok{ IsFib x y × IsFib }\OtherTok{(}\NormalTok{suc x}\OtherTok{)}\NormalTok{ z}\OtherTok{)}

\NormalTok{fib{-}updateSpec }\OtherTok{:}
\NormalTok{    ⇒Dep }\OtherTok{\_}\NormalTok{ FibDep}
         \OtherTok{(}\NormalTok{FreeDep }\OtherTok{\_} \OtherTok{(}\NormalTok{Invariant }\OtherTok{\_}\NormalTok{ FibInvariant}\OtherTok{))}
\NormalTok{         fib{-}update}
\NormalTok{fib{-}updateSpec }\OtherTok{\_} \OtherTok{\_} \OtherTok{=}
\NormalTok{    bindD }\OtherTok{\_} \OtherTok{(λ} \OtherTok{\_} \OtherTok{(}\NormalTok{x , fy , fz}\OtherTok{)} \OtherTok{→}
\NormalTok{        bindD }\OtherTok{(}\NormalTok{suc x , fz , isFibRec fy fz}\OtherTok{)} \OtherTok{(λ} \OtherTok{\_} \OtherTok{\_} \OtherTok{→}
\NormalTok{            returnD }\OtherTok{(\_}\NormalTok{ , fy}\OtherTok{)))}

\NormalTok{fibSpec }\OtherTok{:}\NormalTok{ DepMealy }\OtherTok{\_}\NormalTok{ FibDep fib}
\NormalTok{fibSpec }\OtherTok{=}
\NormalTok{    prog→mealyD fib{-}updateSpec}
                \OtherTok{(}\NormalTok{invariant }\OtherTok{\_} \OtherTok{(}\NormalTok{zero , isFibZero , isFibOne}\OtherTok{))}
\end{Highlighting}
\end{Shaded}

\section{Full Agda Code}\label{full-agda-code}

\begin{Shaded}
\begin{Highlighting}[]
\PreprocessorTok{\{{-}\# OPTIONS {-}{-}without{-}K {-}{-}guardedness \#{-}\}}
\KeywordTok{module}\NormalTok{ cpv }\KeywordTok{where}

\KeywordTok{open} \KeywordTok{import}\NormalTok{ Agda}\OtherTok{.}\NormalTok{Builtin}\OtherTok{.}\NormalTok{Unit}
\KeywordTok{open} \KeywordTok{import}\NormalTok{ Agda}\OtherTok{.}\NormalTok{Builtin}\OtherTok{.}\NormalTok{Sigma}

\CommentTok{{-}{-} binary products}
\OtherTok{\_}\NormalTok{×}\OtherTok{\_} \OtherTok{:} \DataTypeTok{Set} \OtherTok{→} \DataTypeTok{Set} \OtherTok{→} \DataTypeTok{Set}
\NormalTok{A × B }\OtherTok{=}\NormalTok{ Σ A }\OtherTok{(λ} \OtherTok{\_} \OtherTok{→}\NormalTok{ B}\OtherTok{)}

\KeywordTok{infixr} \DecValTok{5} \OtherTok{\_}\NormalTok{×}\OtherTok{\_}

\CommentTok{{-}{-} binary coproducts}
\KeywordTok{data} \OtherTok{\_}\NormalTok{+}\OtherTok{\_} \OtherTok{(}\NormalTok{A }\OtherTok{:} \DataTypeTok{Set}\OtherTok{)} \OtherTok{(}\NormalTok{B }\OtherTok{:} \DataTypeTok{Set}\OtherTok{)} \OtherTok{:} \DataTypeTok{Set} \KeywordTok{where}
\NormalTok{    inl }\OtherTok{:}\NormalTok{ A }\OtherTok{→}\NormalTok{ A + B}
\NormalTok{    inr }\OtherTok{:}\NormalTok{ B }\OtherTok{→}\NormalTok{ A + B}

\CommentTok{{-}{-} empty type}
\KeywordTok{data}\NormalTok{ ⊥ }\OtherTok{:} \DataTypeTok{Set} \KeywordTok{where}

\CommentTok{{-}{-} polynomial functors}
\NormalTok{Poly }\OtherTok{:} \DataTypeTok{Set₁}
\NormalTok{Poly }\OtherTok{=}\NormalTok{ Σ }\DataTypeTok{Set} \OtherTok{(λ}\NormalTok{ A }\OtherTok{→}\NormalTok{ A }\OtherTok{→} \DataTypeTok{Set}\OtherTok{)}

\CommentTok{{-}{-} action of a polynomial on a set}
\OtherTok{\_}\NormalTok{⊙}\OtherTok{\_} \OtherTok{:}\NormalTok{ Poly }\OtherTok{→} \OtherTok{(}\DataTypeTok{Set} \OtherTok{→} \DataTypeTok{Set}\OtherTok{)}
\OtherTok{(}\NormalTok{A , B}\OtherTok{)}\NormalTok{ ⊙ y }\OtherTok{=}\NormalTok{ Σ A }\OtherTok{(λ}\NormalTok{ x }\OtherTok{→}\NormalTok{ B x }\OtherTok{→}\NormalTok{ y}\OtherTok{)}

\CommentTok{{-}{-} polynomial morphisms}
\OtherTok{\_}\NormalTok{⇒}\OtherTok{\_} \OtherTok{:}\NormalTok{ Poly }\OtherTok{→} \OtherTok{(}\DataTypeTok{Set} \OtherTok{→} \DataTypeTok{Set}\OtherTok{)} \OtherTok{→} \DataTypeTok{Set}
\OtherTok{(}\NormalTok{A , B}\OtherTok{)}\NormalTok{ ⇒ F }\OtherTok{=} \OtherTok{(}\NormalTok{x }\OtherTok{:}\NormalTok{ A}\OtherTok{)} \OtherTok{→}\NormalTok{ F }\OtherTok{(}\NormalTok{B x}\OtherTok{)}

\CommentTok{{-}{-} free monad on a polynomial}
\KeywordTok{data}\NormalTok{ Free }\OtherTok{(}\NormalTok{p }\OtherTok{:}\NormalTok{ Poly}\OtherTok{)} \OtherTok{(}\NormalTok{C }\OtherTok{:} \DataTypeTok{Set}\OtherTok{)} \OtherTok{:} \DataTypeTok{Set} \KeywordTok{where}
\NormalTok{    return }\OtherTok{:}\NormalTok{ C }\OtherTok{→}\NormalTok{ Free p C}
\NormalTok{    bind }\OtherTok{:} \OtherTok{(}\NormalTok{x }\OtherTok{:}\NormalTok{ p }\OtherTok{.}\NormalTok{fst}\OtherTok{)} \OtherTok{→} \OtherTok{(}\NormalTok{p }\OtherTok{.}\NormalTok{snd x }\OtherTok{→}\NormalTok{ Free p C}\OtherTok{)} \OtherTok{→}\NormalTok{ Free p C}

\CommentTok{{-}{-} monadic bind/Kleisli lift for Free monads}
\OtherTok{\_}\NormalTok{\textgreater{}\textgreater{}=}\OtherTok{\_} \OtherTok{:} \OtherTok{\{}\NormalTok{p }\OtherTok{:}\NormalTok{ Poly}\OtherTok{\}} \OtherTok{\{}\NormalTok{A B }\OtherTok{:} \DataTypeTok{Set}\OtherTok{\}} 
      \OtherTok{→}\NormalTok{ Free p A }\OtherTok{→} \OtherTok{(}\NormalTok{A }\OtherTok{→}\NormalTok{ Free p B}\OtherTok{)} \OtherTok{→}\NormalTok{ Free p B}
\OtherTok{(}\NormalTok{return a}\OtherTok{)}\NormalTok{ \textgreater{}\textgreater{}= f }\OtherTok{=}\NormalTok{ f a}
\OtherTok{(}\NormalTok{bind x k}\OtherTok{)}\NormalTok{ \textgreater{}\textgreater{}= f }\OtherTok{=}\NormalTok{ bind x }\OtherTok{(λ}\NormalTok{ b }\OtherTok{→}\NormalTok{ k b \textgreater{}\textgreater{}= f}\OtherTok{)}

\KeywordTok{syntax}\NormalTok{ bind a }\OtherTok{(λ}\NormalTok{ b }\OtherTok{→}\NormalTok{ e}\OtherTok{)} \OtherTok{=}\NormalTok{ call[ b ← a ] e}

\NormalTok{\textgreater{}\textgreater{}={-}syntax }\OtherTok{=} \OtherTok{\_}\NormalTok{\textgreater{}\textgreater{}=}\OtherTok{\_}

\KeywordTok{syntax}\NormalTok{ \textgreater{}\textgreater{}={-}syntax m }\OtherTok{(λ}\NormalTok{ b }\OtherTok{→}\NormalTok{ e}\OtherTok{)} \OtherTok{=}\NormalTok{ do[ b ← m ] e}

\CommentTok{{-}{-} sums of polynomials}
\NormalTok{sum }\OtherTok{:} \OtherTok{(}\NormalTok{U }\OtherTok{:} \DataTypeTok{Set}\OtherTok{)} \OtherTok{→} \OtherTok{(}\NormalTok{p }\OtherTok{:}\NormalTok{ U }\OtherTok{→}\NormalTok{ Poly}\OtherTok{)} \OtherTok{→}\NormalTok{ Poly}
\NormalTok{sum U p }\OtherTok{.}\NormalTok{fst }\OtherTok{=}\NormalTok{ Σ U }\OtherTok{(λ}\NormalTok{ u }\OtherTok{→}\NormalTok{ p u }\OtherTok{.}\NormalTok{fst}\OtherTok{)}
\NormalTok{sum U p }\OtherTok{.}\NormalTok{snd }\OtherTok{(}\NormalTok{u , x}\OtherTok{)} \OtherTok{=}\NormalTok{ p u }\OtherTok{.}\NormalTok{snd x}

\CommentTok{{-}{-} binary sum}
\OtherTok{\_}\NormalTok{⊕}\OtherTok{\_} \OtherTok{:}\NormalTok{ Poly }\OtherTok{→}\NormalTok{ Poly }\OtherTok{→}\NormalTok{ Poly}
\OtherTok{((}\NormalTok{A , B}\OtherTok{)}\NormalTok{ ⊕ }\OtherTok{(}\NormalTok{C , D}\OtherTok{))} \OtherTok{.}\NormalTok{fst }\OtherTok{=}\NormalTok{ A + C}
\OtherTok{((}\NormalTok{A , B}\OtherTok{)}\NormalTok{ ⊕ }\OtherTok{(}\NormalTok{C , D}\OtherTok{))} \OtherTok{.}\NormalTok{snd }\OtherTok{(}\NormalTok{inl a}\OtherTok{)} \OtherTok{=}\NormalTok{ B a}
\OtherTok{((}\NormalTok{A , B}\OtherTok{)}\NormalTok{ ⊕ }\OtherTok{(}\NormalTok{C , D}\OtherTok{))} \OtherTok{.}\NormalTok{snd }\OtherTok{(}\NormalTok{inr c}\OtherTok{)} \OtherTok{=}\NormalTok{ D c}

\CommentTok{{-}{-} nullary sum}
\NormalTok{zeroPoly }\OtherTok{:}\NormalTok{ Poly}
\NormalTok{zeroPoly }\OtherTok{.}\NormalTok{fst }\OtherTok{=}\NormalTok{ ⊥}
\NormalTok{zeroPoly }\OtherTok{.}\NormalTok{snd }\OtherTok{()}

\CommentTok{{-}{-} examples: fold, append, concat}
\KeywordTok{module}\NormalTok{ example1 }\KeywordTok{where}

    \CommentTok{{-}{-} simple list type}
    \KeywordTok{data}\NormalTok{ List }\OtherTok{(}\NormalTok{A }\OtherTok{:} \DataTypeTok{Set}\OtherTok{)} \OtherTok{:} \DataTypeTok{Set} \KeywordTok{where}
\NormalTok{        nil  }\OtherTok{:}\NormalTok{ List A}
\NormalTok{        cons }\OtherTok{:}\NormalTok{ A }\OtherTok{→}\NormalTok{ List A }\OtherTok{→}\NormalTok{ List A}

    \CommentTok{{-}{-} labels for fold dependencies}
    \KeywordTok{data}\NormalTok{ foldlabels }\OtherTok{:} \DataTypeTok{Set} \KeywordTok{where}
\NormalTok{        base }\OtherTok{:}\NormalTok{ foldlabels}
\NormalTok{        step }\OtherTok{:}\NormalTok{ foldlabels}

    \CommentTok{{-}{-} interface for base case of a fold}
\NormalTok{    Base }\OtherTok{:} \OtherTok{(}\NormalTok{A C }\OtherTok{:} \DataTypeTok{Set}\OtherTok{)} \OtherTok{→}\NormalTok{ Poly}
\NormalTok{    Base A C }\OtherTok{=} \OtherTok{(}\NormalTok{A , }\OtherTok{λ} \OtherTok{\_} \OtherTok{→}\NormalTok{ C}\OtherTok{)}

    \CommentTok{{-}{-} interface for recursive step of a fold}
\NormalTok{    Step }\OtherTok{:} \OtherTok{(}\NormalTok{A B C }\OtherTok{:} \DataTypeTok{Set}\OtherTok{)} \OtherTok{→}\NormalTok{ Poly}
\NormalTok{    Step A B C }\OtherTok{=} \OtherTok{(}\NormalTok{A × B × C , }\OtherTok{λ} \OtherTok{\_} \OtherTok{→}\NormalTok{ C}\OtherTok{)}

    \CommentTok{{-}{-} interface for a fold}
\NormalTok{    Fold }\OtherTok{:} \OtherTok{(}\NormalTok{A B C }\OtherTok{:} \DataTypeTok{Set}\OtherTok{)} \OtherTok{→}\NormalTok{ Poly}
\NormalTok{    Fold A B C }\OtherTok{=} \OtherTok{(}\NormalTok{A × List B , }\OtherTok{λ} \OtherTok{\_} \OtherTok{→}\NormalTok{ C}\OtherTok{)}

    \CommentTok{{-}{-} fold as a program module}
\NormalTok{    fold }\OtherTok{:} \OtherTok{\{}\NormalTok{A B C }\OtherTok{:} \DataTypeTok{Set}\OtherTok{\}} 
         \OtherTok{→}\NormalTok{ Fold A B C }
\NormalTok{           ⇒ Free }\OtherTok{(}\NormalTok{sum foldlabels }
                       \OtherTok{(λ\{}\NormalTok{ base }\OtherTok{→}\NormalTok{ Base A C }
                         \OtherTok{;}\NormalTok{ step }\OtherTok{→}\NormalTok{ Step A B C}\OtherTok{\}))}
\NormalTok{    fold }\OtherTok{(}\NormalTok{a , nil}\OtherTok{)} \OtherTok{=} 
\NormalTok{        call[ c ← }\OtherTok{(}\NormalTok{base , a}\OtherTok{)}\NormalTok{ ] }
\NormalTok{        return c}
\NormalTok{    fold }\OtherTok{(}\NormalTok{a , cons b bs}\OtherTok{)} \OtherTok{=} 
\NormalTok{        do[ c ← fold }\OtherTok{(}\NormalTok{a , bs}\OtherTok{)}\NormalTok{ ] }
\NormalTok{        call[ c\textquotesingle{} ← }\OtherTok{(}\NormalTok{step , }\OtherTok{(}\NormalTok{a , b , c}\OtherTok{))}\NormalTok{ ] }
\NormalTok{        return c\textquotesingle{}}

    \CommentTok{{-}{-} program modules for append}
\NormalTok{    appendBase }\OtherTok{:} \OtherTok{\{}\NormalTok{A }\OtherTok{:} \DataTypeTok{Set}\OtherTok{\}} 
               \OtherTok{→}\NormalTok{ Base }\OtherTok{(}\NormalTok{List A}\OtherTok{)} \OtherTok{(}\NormalTok{List A}\OtherTok{)}\NormalTok{ ⇒ Free zeroPoly}
\NormalTok{    appendBase xs }\OtherTok{=}\NormalTok{ return xs}

\NormalTok{    appendStep }\OtherTok{:} \OtherTok{\{}\NormalTok{A }\OtherTok{:} \DataTypeTok{Set}\OtherTok{\}} 
               \OtherTok{→}\NormalTok{ Step }\OtherTok{(}\NormalTok{List A}\OtherTok{)}\NormalTok{ A }\OtherTok{(}\NormalTok{List A}\OtherTok{)}\NormalTok{ ⇒ Free zeroPoly}
\NormalTok{    appendStep }\OtherTok{(\_}\NormalTok{ , x , xs}\OtherTok{)} \OtherTok{=}\NormalTok{ return }\OtherTok{(}\NormalTok{cons x xs}\OtherTok{)}

    \CommentTok{{-}{-} ...and for concat}
\NormalTok{    concatBase }\OtherTok{:} \OtherTok{\{}\NormalTok{A }\OtherTok{:} \DataTypeTok{Set}\OtherTok{\}} 
               \OtherTok{→}\NormalTok{ Base ⊤ }\OtherTok{(}\NormalTok{List A}\OtherTok{)} 
\NormalTok{                 ⇒ Free zeroPoly}
\NormalTok{    concatBase }\OtherTok{\_} \OtherTok{=}\NormalTok{ return nil}

\NormalTok{    concatStep }\OtherTok{:} \OtherTok{\{}\NormalTok{A }\OtherTok{:} \DataTypeTok{Set}\OtherTok{\}} 
               \OtherTok{→}\NormalTok{ Step ⊤ }\OtherTok{(}\NormalTok{List A}\OtherTok{)} \OtherTok{(}\NormalTok{List A}\OtherTok{)} 
\NormalTok{                 ⇒ Free }\OtherTok{(}\NormalTok{Fold }\OtherTok{(}\NormalTok{List A}\OtherTok{)}\NormalTok{ A }\OtherTok{(}\NormalTok{List A}\OtherTok{))}
\NormalTok{    concatStep }\OtherTok{(\_}\NormalTok{ , xs , ys}\OtherTok{)} \OtherTok{=}
\NormalTok{        call[ zs ← }\OtherTok{(}\NormalTok{ys , xs}\OtherTok{)}\NormalTok{ ] }
\NormalTok{        return zs}

\CommentTok{{-}{-} mapping a polynomial kleisli morphism over a free monad}
\NormalTok{Free⇒ }\OtherTok{:} \OtherTok{\{}\NormalTok{p q }\OtherTok{:}\NormalTok{ Poly}\OtherTok{\}} \OtherTok{\{}\NormalTok{E }\OtherTok{:} \DataTypeTok{Set}\OtherTok{\}} 
      \OtherTok{→}\NormalTok{ Free p E }\OtherTok{→} \OtherTok{(}\NormalTok{p ⇒ Free q}\OtherTok{)} \OtherTok{→}\NormalTok{ Free q E}
\NormalTok{Free⇒ }\OtherTok{(}\NormalTok{return e}\OtherTok{)}\NormalTok{ f }\OtherTok{=}\NormalTok{ return e}
\NormalTok{Free⇒ }\OtherTok{(}\NormalTok{bind a k}\OtherTok{)}\NormalTok{ f }\OtherTok{=} \OtherTok{(}\NormalTok{f a}\OtherTok{)}\NormalTok{ \textgreater{}\textgreater{}= }\OtherTok{(λ}\NormalTok{ b }\OtherTok{→}\NormalTok{ Free⇒ }\OtherTok{(}\NormalTok{k b}\OtherTok{)}\NormalTok{ f}\OtherTok{)}

\CommentTok{{-}{-} Kleisli composition for the free{-}monad{-}monad}
\OtherTok{\_}\NormalTok{∘}\OtherTok{\_} \OtherTok{:} \OtherTok{\{}\NormalTok{p q r }\OtherTok{:}\NormalTok{ Poly}\OtherTok{\}} 
    \OtherTok{→} \OtherTok{(}\NormalTok{p ⇒ Free q}\OtherTok{)} \OtherTok{→} \OtherTok{(}\NormalTok{q ⇒ Free r}\OtherTok{)} \OtherTok{→}\NormalTok{ p ⇒ Free r}
\NormalTok{f ∘ g }\OtherTok{=} \OtherTok{λ}\NormalTok{ a }\OtherTok{→}\NormalTok{ Free⇒ }\OtherTok{(}\NormalTok{f a}\OtherTok{)}\NormalTok{ g}

\CommentTok{{-}{-} U{-}ary composition}
\NormalTok{comp }\OtherTok{:} \OtherTok{(}\NormalTok{U }\OtherTok{:} \DataTypeTok{Set}\OtherTok{)} \OtherTok{\{}\NormalTok{p }\OtherTok{:}\NormalTok{ Poly}\OtherTok{\}} \OtherTok{\{}\NormalTok{q }\OtherTok{:}\NormalTok{ U }\OtherTok{→}\NormalTok{ Poly}\OtherTok{\}} \OtherTok{\{}\NormalTok{r }\OtherTok{:}\NormalTok{ Poly}\OtherTok{\}} 
     \OtherTok{→} \OtherTok{(}\NormalTok{p ⇒ Free }\OtherTok{(}\NormalTok{sum U q}\OtherTok{))} 
     \OtherTok{→} \OtherTok{((}\NormalTok{u }\OtherTok{:}\NormalTok{ U}\OtherTok{)} \OtherTok{→}\NormalTok{ q u ⇒ Free r}\OtherTok{)} 
     \OtherTok{→}\NormalTok{ p ⇒ Free r}
\NormalTok{comp U }\OtherTok{\{}\NormalTok{p }\OtherTok{=}\NormalTok{ p}\OtherTok{\}} \OtherTok{\{}\NormalTok{q }\OtherTok{=}\NormalTok{ q}\OtherTok{\}} \OtherTok{\{}\NormalTok{r }\OtherTok{=}\NormalTok{ r}\OtherTok{\}}\NormalTok{ f g }\OtherTok{=}\NormalTok{ f ∘ }\OtherTok{(λ} \OtherTok{(}\NormalTok{u , x}\OtherTok{)} \OtherTok{→}\NormalTok{ g u x}\OtherTok{)}

\CommentTok{{-}{-} composing the implementations for append and concat}
\KeywordTok{module}\NormalTok{ example2 }\KeywordTok{where}

    \KeywordTok{open}\NormalTok{ example1}

\NormalTok{    append }\OtherTok{:} \OtherTok{\{}\NormalTok{A }\OtherTok{:} \DataTypeTok{Set}\OtherTok{\}} 
           \OtherTok{→}\NormalTok{ Fold }\OtherTok{(}\NormalTok{List A}\OtherTok{)}\NormalTok{ A }\OtherTok{(}\NormalTok{List A}\OtherTok{)}\NormalTok{ ⇒ Free zeroPoly}
\NormalTok{    append }\OtherTok{=} 
\NormalTok{        comp foldlabels fold }
             \OtherTok{(λ\{}\NormalTok{ base }\OtherTok{→}\NormalTok{ appendBase }
               \OtherTok{;}\NormalTok{ step }\OtherTok{→}\NormalTok{ appendStep}\OtherTok{\})}

\NormalTok{    concat }\OtherTok{:} \OtherTok{\{}\NormalTok{A }\OtherTok{:} \DataTypeTok{Set}\OtherTok{\}} 
           \OtherTok{→}\NormalTok{ Fold ⊤ }\OtherTok{(}\NormalTok{List A}\OtherTok{)} \OtherTok{(}\NormalTok{List A}\OtherTok{)}\NormalTok{ ⇒ Free zeroPoly}
\NormalTok{    concat }\OtherTok{=} 
\NormalTok{        comp foldlabels fold }
             \OtherTok{(λ\{}\NormalTok{ base }\OtherTok{→}\NormalTok{ concatBase }
               \OtherTok{;}\NormalTok{ step }\OtherTok{→}\NormalTok{ concatStep ∘ append}\OtherTok{\})}

\CommentTok{{-}{-} wiring diagrams}
\KeywordTok{module}\NormalTok{ WD }\OtherTok{(}\NormalTok{Boxes }\OtherTok{:} \DataTypeTok{Set}\OtherTok{)}
          \OtherTok{(}\NormalTok{Arity }\OtherTok{:}\NormalTok{ Boxes }\OtherTok{→} \DataTypeTok{Set}\OtherTok{)}
          \OtherTok{(}\NormalTok{Dom }\OtherTok{:} \OtherTok{(}\NormalTok{b }\OtherTok{:}\NormalTok{ Boxes}\OtherTok{)} \OtherTok{→}\NormalTok{ Arity b }\OtherTok{→}\NormalTok{ Poly}\OtherTok{)}
          \OtherTok{(}\NormalTok{Cod }\OtherTok{:}\NormalTok{ Boxes }\OtherTok{→}\NormalTok{ Poly}\OtherTok{)}
          \OtherTok{(}\NormalTok{Inputs }\OtherTok{:} \DataTypeTok{Set}\OtherTok{)} \OtherTok{(}\NormalTok{inputs }\OtherTok{:}\NormalTok{ Inputs }\OtherTok{→}\NormalTok{ Poly}\OtherTok{)} \KeywordTok{where}

    \KeywordTok{data}\NormalTok{ Wiring }\OtherTok{:} \OtherTok{(}\NormalTok{output }\OtherTok{:}\NormalTok{ Poly}\OtherTok{)} \OtherTok{→} \DataTypeTok{Set₁} \KeywordTok{where}
\NormalTok{        wire }\OtherTok{:} \OtherTok{(}\NormalTok{i }\OtherTok{:}\NormalTok{ Inputs}\OtherTok{)} \OtherTok{→}\NormalTok{ Wiring }\OtherTok{(}\NormalTok{inputs i}\OtherTok{)}
\NormalTok{        box }\OtherTok{:} \OtherTok{(}\NormalTok{b }\OtherTok{:}\NormalTok{ Boxes}\OtherTok{)} \OtherTok{→} \OtherTok{((}\NormalTok{a }\OtherTok{:}\NormalTok{ Arity b}\OtherTok{)} \OtherTok{→}\NormalTok{ Wiring }\OtherTok{(}\NormalTok{Dom b a}\OtherTok{))} 
            \OtherTok{→}\NormalTok{ Wiring }\OtherTok{(}\NormalTok{Cod b}\OtherTok{)}

\KeywordTok{open}\NormalTok{ WD }\KeywordTok{public}

\NormalTok{compose }\OtherTok{:} \OtherTok{\{}\NormalTok{Boxes }\OtherTok{:} \DataTypeTok{Set}\OtherTok{\}} \OtherTok{\{}\NormalTok{Arity }\OtherTok{:}\NormalTok{ Boxes }\OtherTok{→} \DataTypeTok{Set}\OtherTok{\}} 
        \OtherTok{→} \OtherTok{\{}\NormalTok{Dom }\OtherTok{:} \OtherTok{(}\NormalTok{b }\OtherTok{:}\NormalTok{ Boxes}\OtherTok{)} \OtherTok{→}\NormalTok{ Arity b }\OtherTok{→}\NormalTok{ Poly}\OtherTok{\}} 
        \OtherTok{→} \OtherTok{\{}\NormalTok{Cod }\OtherTok{:}\NormalTok{ Boxes }\OtherTok{→}\NormalTok{ Poly}\OtherTok{\}} \OtherTok{\{}\NormalTok{Inputs }\OtherTok{:} \DataTypeTok{Set}\OtherTok{\}} 
        \OtherTok{→} \OtherTok{\{}\NormalTok{inputs }\OtherTok{:}\NormalTok{ Inputs }\OtherTok{→}\NormalTok{ Poly}\OtherTok{\}} \OtherTok{\{}\NormalTok{output }\OtherTok{:}\NormalTok{ Poly}\OtherTok{\}}
        \OtherTok{→}\NormalTok{ Wiring Boxes Arity Dom Cod Inputs inputs output}
        \OtherTok{→} \OtherTok{((}\NormalTok{b }\OtherTok{:}\NormalTok{ Boxes}\OtherTok{)} \OtherTok{→}\NormalTok{ Cod b ⇒ Free }\OtherTok{(}\NormalTok{sum }\OtherTok{(}\NormalTok{Arity b}\OtherTok{)} \OtherTok{(}\NormalTok{Dom b}\OtherTok{)))}
        \OtherTok{→}\NormalTok{ output ⇒ Free }\OtherTok{(}\NormalTok{sum Inputs inputs}\OtherTok{)}
\NormalTok{compose }\OtherTok{(}\NormalTok{wire i}\OtherTok{)}\NormalTok{ f x }\OtherTok{=}\NormalTok{ bind }\OtherTok{(}\NormalTok{i , x}\OtherTok{)}\NormalTok{ return}
\NormalTok{compose }\OtherTok{(}\NormalTok{box b f}\OtherTok{)}\NormalTok{ g }\OtherTok{=}\NormalTok{ comp }\OtherTok{\_} \OtherTok{(}\NormalTok{g b}\OtherTok{)} \OtherTok{(λ}\NormalTok{ a }\OtherTok{→}\NormalTok{ compose }\OtherTok{(}\NormalTok{f a}\OtherTok{)}\NormalTok{ g}\OtherTok{)}

\CommentTok{{-}{-} Mealy machines}
\KeywordTok{record}\NormalTok{ Mealy }\OtherTok{(}\NormalTok{p }\OtherTok{:}\NormalTok{ Poly}\OtherTok{)} \OtherTok{:} \DataTypeTok{Set} \KeywordTok{where}
    \KeywordTok{coinductive}
    \KeywordTok{field}
\NormalTok{        app }\OtherTok{:} \OtherTok{(}\NormalTok{x }\OtherTok{:}\NormalTok{ p }\OtherTok{.}\NormalTok{fst}\OtherTok{)} \OtherTok{→}\NormalTok{ p }\OtherTok{.}\NormalTok{snd x × Mealy p}

\KeywordTok{open}\NormalTok{ Mealy }\KeywordTok{public}

\CommentTok{{-}{-} run a mealy machine on a "program" of type Free p C}
\NormalTok{run{-}mealy }\OtherTok{:} \OtherTok{\{}\NormalTok{p }\OtherTok{:}\NormalTok{ Poly}\OtherTok{\}} \OtherTok{\{}\NormalTok{C }\OtherTok{:} \DataTypeTok{Set}\OtherTok{\}}
          \OtherTok{→}\NormalTok{ Free p C }\OtherTok{→}\NormalTok{ Mealy p }\OtherTok{→}\NormalTok{ C × Mealy p}
\NormalTok{run{-}mealy }\OtherTok{(}\NormalTok{return c}\OtherTok{)}\NormalTok{ m }\OtherTok{=} \OtherTok{(}\NormalTok{c , m}\OtherTok{)}
\NormalTok{run{-}mealy }\OtherTok{(}\NormalTok{bind x f}\OtherTok{)}\NormalTok{ m }\OtherTok{=}
    \KeywordTok{let} \OtherTok{(}\NormalTok{b , m\textquotesingle{}}\OtherTok{)} \OtherTok{=}\NormalTok{ m }\OtherTok{.}\NormalTok{app x }\KeywordTok{in}
\NormalTok{    run{-}mealy }\OtherTok{(}\NormalTok{f b}\OtherTok{)}\NormalTok{ m\textquotesingle{}}

\CommentTok{{-}{-} use a polynomial morphism p ⇒ Free q to convert a Mealy }
\CommentTok{{-}{-} machine implementing q to a Mealy machine implementing p}
\NormalTok{prog→mealy }\OtherTok{:} \OtherTok{\{}\NormalTok{p q }\OtherTok{:}\NormalTok{ Poly}\OtherTok{\}} 
           \OtherTok{→} \OtherTok{(}\NormalTok{f }\OtherTok{:}\NormalTok{ p ⇒ Free q}\OtherTok{)} \OtherTok{→}\NormalTok{ Mealy q }\OtherTok{→}\NormalTok{ Mealy p}
\NormalTok{prog→mealy }\OtherTok{\{}\NormalTok{p }\OtherTok{=}\NormalTok{ p}\OtherTok{\}} \OtherTok{\{}\NormalTok{q }\OtherTok{=}\NormalTok{ q}\OtherTok{\}}\NormalTok{ f m }\OtherTok{.}\NormalTok{app a }\OtherTok{=}
    \KeywordTok{let} \OtherTok{(}\NormalTok{b , m\textquotesingle{}}\OtherTok{)} \OtherTok{=}\NormalTok{ run{-}mealy }\OtherTok{(}\NormalTok{f a}\OtherTok{)}\NormalTok{ m }\KeywordTok{in} 
    \OtherTok{(}\NormalTok{b , prog→mealy f m\textquotesingle{}}\OtherTok{)}

\CommentTok{{-}{-} type of inputs for the state effect}
\KeywordTok{data}\NormalTok{ StateI }\OtherTok{(}\NormalTok{S }\OtherTok{:} \DataTypeTok{Set}\OtherTok{)} \OtherTok{:} \DataTypeTok{Set} \KeywordTok{where}
\NormalTok{    get }\OtherTok{:}\NormalTok{ StateI S}
\NormalTok{    put }\OtherTok{:}\NormalTok{ S }\OtherTok{→}\NormalTok{ StateI S}

\NormalTok{State }\OtherTok{:} \DataTypeTok{Set} \OtherTok{→}\NormalTok{ Poly}
\NormalTok{State S }\OtherTok{.}\NormalTok{fst }\OtherTok{=}\NormalTok{ StateI S}
\NormalTok{State S }\OtherTok{.}\NormalTok{snd get }\OtherTok{=}\NormalTok{ S}
\NormalTok{State S }\OtherTok{.}\NormalTok{snd }\OtherTok{(}\NormalTok{put s}\OtherTok{)} \OtherTok{=}\NormalTok{ ⊤}

\NormalTok{state{-}mealy }\OtherTok{:} \OtherTok{\{}\NormalTok{S }\OtherTok{:} \DataTypeTok{Set}\OtherTok{\}} \OtherTok{→}\NormalTok{ S }\OtherTok{→}\NormalTok{ Mealy }\OtherTok{(}\NormalTok{State S}\OtherTok{)}
\NormalTok{state{-}mealy s }\OtherTok{.}\NormalTok{app get }\OtherTok{=} \OtherTok{(}\NormalTok{s , state{-}mealy s}\OtherTok{)}
\NormalTok{state{-}mealy s }\OtherTok{.}\NormalTok{app }\OtherTok{(}\NormalTok{put s\textquotesingle{}}\OtherTok{)} \OtherTok{=} \OtherTok{(\_}\NormalTok{ , state{-}mealy s\textquotesingle{}}\OtherTok{)}

\CommentTok{{-}{-} closure of Mealy machines under binary sum}
\OtherTok{\_}\NormalTok{mealy⊕}\OtherTok{\_} \OtherTok{:} \OtherTok{\{}\NormalTok{p q }\OtherTok{:}\NormalTok{ Poly}\OtherTok{\}} 
         \OtherTok{→}\NormalTok{ Mealy p }\OtherTok{→}\NormalTok{ Mealy q }\OtherTok{→}\NormalTok{ Mealy }\OtherTok{(}\NormalTok{p ⊕ q}\OtherTok{)}
\OtherTok{\_}\NormalTok{mealy⊕}\OtherTok{\_}\NormalTok{ m₁ m₂ }\OtherTok{.}\NormalTok{app }\OtherTok{(}\NormalTok{inl a}\OtherTok{)} \OtherTok{=} 
    \KeywordTok{let} \OtherTok{(}\NormalTok{b , m\textquotesingle{}}\OtherTok{)} \OtherTok{=}\NormalTok{ m₁ }\OtherTok{.}\NormalTok{app a }\KeywordTok{in} \OtherTok{(}\NormalTok{b , }\OtherTok{(}\NormalTok{m\textquotesingle{} mealy⊕ m₂}\OtherTok{))}
\OtherTok{\_}\NormalTok{mealy⊕}\OtherTok{\_}\NormalTok{ m₁ m₂ }\OtherTok{.}\NormalTok{app }\OtherTok{(}\NormalTok{inr c}\OtherTok{)} \OtherTok{=} 
    \KeywordTok{let} \OtherTok{(}\NormalTok{d , m\textquotesingle{}}\OtherTok{)} \OtherTok{=}\NormalTok{ m₂ }\OtherTok{.}\NormalTok{app c }\KeywordTok{in} \OtherTok{(}\NormalTok{d , }\OtherTok{(}\NormalTok{m₁ mealy⊕ m\textquotesingle{}}\OtherTok{))}

\CommentTok{{-}{-} closure of Mealy machines under nullary sum}
\NormalTok{mealyzero }\OtherTok{:}\NormalTok{ Mealy zeroPoly}
\NormalTok{mealyzero }\OtherTok{.}\NormalTok{app }\OtherTok{()}

\CommentTok{{-}{-} example: fibonacci}
\KeywordTok{module}\NormalTok{ example4 }\KeywordTok{where}

    \KeywordTok{open} \KeywordTok{import}\NormalTok{ Agda}\OtherTok{.}\NormalTok{Builtin}\OtherTok{.}\NormalTok{Nat}

\NormalTok{    fib{-}update }\OtherTok{:} \OtherTok{(}\NormalTok{⊤ , }\OtherTok{λ} \OtherTok{\_} \OtherTok{→}\NormalTok{ Nat}\OtherTok{)}\NormalTok{ ⇒ Free }\OtherTok{(}\NormalTok{State }\OtherTok{(}\NormalTok{Nat × Nat}\OtherTok{))}
\NormalTok{    fib{-}update }\OtherTok{\_} \OtherTok{=} 
\NormalTok{        call[ }\OtherTok{(}\NormalTok{x , y}\OtherTok{)}\NormalTok{ ← get ] }
\NormalTok{        call[ }\OtherTok{\_}\NormalTok{ ← put }\OtherTok{(}\NormalTok{y , x + y}\OtherTok{)}\NormalTok{ ] }
\NormalTok{        return x}

\NormalTok{    fib }\OtherTok{:}\NormalTok{ Mealy }\OtherTok{(}\NormalTok{⊤ , }\OtherTok{λ} \OtherTok{\_} \OtherTok{→}\NormalTok{ Nat}\OtherTok{)}
\NormalTok{    fib }\OtherTok{=}\NormalTok{ prog→mealy fib{-}update }\OtherTok{(}\NormalTok{state{-}mealy }\OtherTok{(}\DecValTok{0}\NormalTok{ , }\DecValTok{1}\OtherTok{))}

\CommentTok{{-}{-} dependent polynomials}
\NormalTok{DepPoly }\OtherTok{:}\NormalTok{ Poly }\OtherTok{→} \DataTypeTok{Set₁}
\NormalTok{DepPoly }\OtherTok{(}\NormalTok{A , B}\OtherTok{)} \OtherTok{=}\NormalTok{ Σ }\OtherTok{(}\NormalTok{A }\OtherTok{→} \DataTypeTok{Set}\OtherTok{)} \OtherTok{(λ}\NormalTok{ C }\OtherTok{→} \OtherTok{(}\NormalTok{x }\OtherTok{:}\NormalTok{ A}\OtherTok{)} \OtherTok{→}\NormalTok{ C x }\OtherTok{→}\NormalTok{ B x }\OtherTok{→} \DataTypeTok{Set}\OtherTok{)}

\CommentTok{{-}{-} action of a dependent polynomial}
\NormalTok{⊙Dep }\OtherTok{:} \OtherTok{(}\NormalTok{p }\OtherTok{:}\NormalTok{ Poly}\OtherTok{)} \OtherTok{→}\NormalTok{ DepPoly p }\OtherTok{→} \OtherTok{(}\NormalTok{E }\OtherTok{:} \DataTypeTok{Set}\OtherTok{)} \OtherTok{→} \OtherTok{(}\NormalTok{E }\OtherTok{→} \DataTypeTok{Set}\OtherTok{)} \OtherTok{→}\NormalTok{ p ⊙ E }\OtherTok{→} \DataTypeTok{Set}
\NormalTok{⊙Dep }\OtherTok{(}\NormalTok{A , B}\OtherTok{)} \OtherTok{(}\NormalTok{C , D}\OtherTok{)}\NormalTok{ E F }\OtherTok{(}\NormalTok{a , f}\OtherTok{)} \OtherTok{=} 
\NormalTok{    Σ }\OtherTok{(}\NormalTok{C a}\OtherTok{)} \OtherTok{(λ}\NormalTok{ c }\OtherTok{→} \OtherTok{(}\NormalTok{b }\OtherTok{:}\NormalTok{ B a}\OtherTok{)} \OtherTok{→}\NormalTok{ D a c b }\OtherTok{→}\NormalTok{ F }\OtherTok{(}\NormalTok{f b}\OtherTok{))}

\CommentTok{{-}{-} morphisms of dependent polynomials}
\NormalTok{⇒Dep }\OtherTok{:} \OtherTok{(}\NormalTok{p }\OtherTok{:}\NormalTok{ Poly}\OtherTok{)} \OtherTok{→}\NormalTok{ DepPoly p }\OtherTok{→} \OtherTok{\{}\NormalTok{F }\OtherTok{:} \DataTypeTok{Set} \OtherTok{→} \DataTypeTok{Set}\OtherTok{\}}
     \OtherTok{→} \OtherTok{((}\NormalTok{X }\OtherTok{:} \DataTypeTok{Set}\OtherTok{)} \OtherTok{→} \OtherTok{(}\NormalTok{X }\OtherTok{→} \DataTypeTok{Set}\OtherTok{)} \OtherTok{→}\NormalTok{ F X }\OtherTok{→} \DataTypeTok{Set}\OtherTok{)} 
     \OtherTok{→} \OtherTok{(}\NormalTok{f }\OtherTok{:}\NormalTok{ p ⇒ F}\OtherTok{)} \OtherTok{→} \DataTypeTok{Set}
\NormalTok{⇒Dep }\OtherTok{(}\NormalTok{A , B}\OtherTok{)} \OtherTok{(}\NormalTok{C , D}\OtherTok{)}\NormalTok{ G f }\OtherTok{=}
    \OtherTok{(}\NormalTok{a }\OtherTok{:}\NormalTok{ A}\OtherTok{)} \OtherTok{(}\NormalTok{c }\OtherTok{:}\NormalTok{ C a}\OtherTok{)} \OtherTok{→}\NormalTok{ G }\OtherTok{(}\NormalTok{B a}\OtherTok{)} \OtherTok{(}\NormalTok{D a c}\OtherTok{)} \OtherTok{(}\NormalTok{f a}\OtherTok{)}

\CommentTok{{-}{-} free dependent monads}
\KeywordTok{data}\NormalTok{ FreeDep }\OtherTok{(}\NormalTok{p }\OtherTok{:}\NormalTok{ Poly}\OtherTok{)} \OtherTok{(}\NormalTok{r }\OtherTok{:}\NormalTok{ DepPoly p}\OtherTok{)} 
             \OtherTok{(}\NormalTok{E }\OtherTok{:} \DataTypeTok{Set}\OtherTok{)} \OtherTok{(}\NormalTok{F }\OtherTok{:}\NormalTok{ E }\OtherTok{→} \DataTypeTok{Set}\OtherTok{)} \OtherTok{:}\NormalTok{ Free p E }\OtherTok{→} \DataTypeTok{Set} \KeywordTok{where}
\NormalTok{    returnD }\OtherTok{:} \OtherTok{\{}\NormalTok{e }\OtherTok{:}\NormalTok{ E}\OtherTok{\}} \OtherTok{→}\NormalTok{ F e }\OtherTok{→}\NormalTok{ FreeDep p r E F }\OtherTok{(}\NormalTok{return e}\OtherTok{)}
\NormalTok{    bindD }\OtherTok{:} \OtherTok{\{}\NormalTok{a }\OtherTok{:}\NormalTok{ p }\OtherTok{.}\NormalTok{fst}\OtherTok{\}} \OtherTok{(}\NormalTok{c }\OtherTok{:}\NormalTok{ r }\OtherTok{.}\NormalTok{fst a}\OtherTok{)} 
          \OtherTok{→} \OtherTok{\{}\NormalTok{k }\OtherTok{:} \OtherTok{(}\NormalTok{p }\OtherTok{.}\NormalTok{snd a}\OtherTok{)} \OtherTok{→}\NormalTok{ Free p E}\OtherTok{\}}
          \OtherTok{→} \OtherTok{((}\NormalTok{b }\OtherTok{:}\NormalTok{ p }\OtherTok{.}\NormalTok{snd a}\OtherTok{)} \OtherTok{→}\NormalTok{ r }\OtherTok{.}\NormalTok{snd a c b }
                            \OtherTok{→}\NormalTok{ FreeDep p r E F }\OtherTok{(}\NormalTok{k b}\OtherTok{))}
          \OtherTok{→}\NormalTok{ FreeDep p r E F }\OtherTok{(}\NormalTok{bind a k}\OtherTok{)}

\CommentTok{{-}{-} monadic bind for FreeDep}
\NormalTok{\textgreater{}\textgreater{}=Dep }\OtherTok{:} \OtherTok{\{}\NormalTok{p }\OtherTok{:}\NormalTok{ Poly}\OtherTok{\}} \OtherTok{\{}\NormalTok{r }\OtherTok{:}\NormalTok{ DepPoly p}\OtherTok{\}} 
       \OtherTok{→} \OtherTok{\{}\NormalTok{E }\OtherTok{:} \DataTypeTok{Set}\OtherTok{\}} \OtherTok{\{}\NormalTok{F }\OtherTok{:}\NormalTok{ E }\OtherTok{→} \DataTypeTok{Set}\OtherTok{\}} \OtherTok{\{}\NormalTok{G }\OtherTok{:} \DataTypeTok{Set}\OtherTok{\}} \OtherTok{\{}\NormalTok{H }\OtherTok{:}\NormalTok{ G }\OtherTok{→} \DataTypeTok{Set}\OtherTok{\}}
       \OtherTok{→} \OtherTok{(}\NormalTok{f }\OtherTok{:}\NormalTok{ Free p E}\OtherTok{)} \OtherTok{→}\NormalTok{ FreeDep p r E F f}
       \OtherTok{→} \OtherTok{\{}\NormalTok{g }\OtherTok{:}\NormalTok{ E }\OtherTok{→}\NormalTok{ Free p G}\OtherTok{\}} 
       \OtherTok{→} \OtherTok{((}\NormalTok{e }\OtherTok{:}\NormalTok{ E}\OtherTok{)} \OtherTok{→}\NormalTok{ F e }\OtherTok{→}\NormalTok{ FreeDep p r G H }\OtherTok{(}\NormalTok{g e}\OtherTok{))}
       \OtherTok{→}\NormalTok{ FreeDep p r G H }\OtherTok{(}\NormalTok{f \textgreater{}\textgreater{}= g}\OtherTok{)}
\NormalTok{\textgreater{}\textgreater{}=Dep }\OtherTok{(}\NormalTok{return e}\OtherTok{)} \OtherTok{(}\NormalTok{returnD f}\OtherTok{)}\NormalTok{ gg }\OtherTok{=}\NormalTok{ gg e f}
\NormalTok{\textgreater{}\textgreater{}=Dep }\OtherTok{(}\NormalTok{bind a k}\OtherTok{)} \OtherTok{(}\NormalTok{bindD c h}\OtherTok{)}\NormalTok{ gg }\OtherTok{=} 
\NormalTok{    bindD c }\OtherTok{(λ}\NormalTok{ b x }\OtherTok{→}\NormalTok{ \textgreater{}\textgreater{}=Dep }\OtherTok{(}\NormalTok{k b}\OtherTok{)} \OtherTok{(}\NormalTok{h b x}\OtherTok{)}\NormalTok{ gg}\OtherTok{)}

\CommentTok{{-}{-} sums of dependent polynomials}
\NormalTok{sumDep }\OtherTok{:} \OtherTok{(}\NormalTok{U }\OtherTok{:} \DataTypeTok{Set}\OtherTok{)} \OtherTok{(}\NormalTok{p }\OtherTok{:}\NormalTok{ U }\OtherTok{→}\NormalTok{ Poly}\OtherTok{)} 
       \OtherTok{→} \OtherTok{(}\NormalTok{r }\OtherTok{:} \OtherTok{(}\NormalTok{u }\OtherTok{:}\NormalTok{ U}\OtherTok{)} \OtherTok{→}\NormalTok{ DepPoly }\OtherTok{(}\NormalTok{p u}\OtherTok{))} \OtherTok{→}\NormalTok{ DepPoly }\OtherTok{(}\NormalTok{sum U p}\OtherTok{)}
\NormalTok{sumDep U p r }\OtherTok{.}\NormalTok{fst }\OtherTok{(}\NormalTok{u , x}\OtherTok{)} \OtherTok{=}\NormalTok{ r u }\OtherTok{.}\NormalTok{fst x}
\NormalTok{sumDep U p r }\OtherTok{.}\NormalTok{snd }\OtherTok{(}\NormalTok{u , x}\OtherTok{)}\NormalTok{ y }\OtherTok{=}\NormalTok{ r u }\OtherTok{.}\NormalTok{snd x y}

\CommentTok{{-}{-} binary coproduct of dependent polynomials}
\NormalTok{depPoly⊕ }\OtherTok{:} \OtherTok{\{}\NormalTok{p q }\OtherTok{:}\NormalTok{ Poly}\OtherTok{\}} \OtherTok{(}\NormalTok{r }\OtherTok{:}\NormalTok{ DepPoly p}\OtherTok{)} \OtherTok{(}\NormalTok{s }\OtherTok{:}\NormalTok{ DepPoly q}\OtherTok{)}
         \OtherTok{→}\NormalTok{ DepPoly }\OtherTok{(}\NormalTok{p ⊕ q}\OtherTok{)}
\NormalTok{depPoly⊕ }\OtherTok{(}\NormalTok{A , B}\OtherTok{)} \OtherTok{(}\NormalTok{C , D}\OtherTok{)} \OtherTok{.}\NormalTok{fst }\OtherTok{(}\NormalTok{inl a}\OtherTok{)} \OtherTok{=}\NormalTok{ A a}
\NormalTok{depPoly⊕ }\OtherTok{(}\NormalTok{A , B}\OtherTok{)} \OtherTok{(}\NormalTok{C , D}\OtherTok{)} \OtherTok{.}\NormalTok{fst }\OtherTok{(}\NormalTok{inr c}\OtherTok{)} \OtherTok{=}\NormalTok{ C c}
\NormalTok{depPoly⊕ }\OtherTok{(}\NormalTok{A , B}\OtherTok{)} \OtherTok{(}\NormalTok{C , D}\OtherTok{)} \OtherTok{.}\NormalTok{snd }\OtherTok{(}\NormalTok{inl a}\OtherTok{)}\NormalTok{ y }\OtherTok{=}\NormalTok{ B a y}
\NormalTok{depPoly⊕ }\OtherTok{(}\NormalTok{A , B}\OtherTok{)} \OtherTok{(}\NormalTok{C , D}\OtherTok{)} \OtherTok{.}\NormalTok{snd }\OtherTok{(}\NormalTok{inr c}\OtherTok{)}\NormalTok{ y }\OtherTok{=}\NormalTok{ D c y}

\CommentTok{{-}{-} nullary coproduct of dependent polynomials}
\NormalTok{depPolyzero }\OtherTok{:}\NormalTok{ DepPoly zeroPoly}
\NormalTok{depPolyzero }\OtherTok{.}\NormalTok{fst }\OtherTok{()}
\NormalTok{depPolyzero }\OtherTok{.}\NormalTok{snd }\OtherTok{()}

\CommentTok{{-}{-} mapping a dependent polynomial morphism over }
\CommentTok{{-}{-} a free dependent monad}
\NormalTok{FreeDep⇒ }\OtherTok{:} \OtherTok{\{}\NormalTok{p q }\OtherTok{:}\NormalTok{ Poly}\OtherTok{\}} \OtherTok{\{}\NormalTok{r }\OtherTok{:}\NormalTok{ DepPoly p}\OtherTok{\}} \OtherTok{\{}\NormalTok{s }\OtherTok{:}\NormalTok{ DepPoly q}\OtherTok{\}} 
         \OtherTok{→} \OtherTok{\{}\NormalTok{E }\OtherTok{:} \DataTypeTok{Set}\OtherTok{\}} \OtherTok{\{}\NormalTok{F }\OtherTok{:}\NormalTok{ E }\OtherTok{→} \DataTypeTok{Set}\OtherTok{\}}
         \OtherTok{→} \OtherTok{(}\NormalTok{f }\OtherTok{:}\NormalTok{ Free p E}\OtherTok{)} \OtherTok{→}\NormalTok{ FreeDep p r E F f}
         \OtherTok{→} \OtherTok{(}\NormalTok{g }\OtherTok{:}\NormalTok{ p ⇒ Free q}\OtherTok{)} \OtherTok{→} \OtherTok{(}\NormalTok{⇒Dep p r }\OtherTok{(}\NormalTok{FreeDep q s}\OtherTok{)}\NormalTok{ g}\OtherTok{)}
         \OtherTok{→}\NormalTok{ FreeDep q s E F }\OtherTok{(}\NormalTok{Free⇒ f g}\OtherTok{)}
\NormalTok{FreeDep⇒ }\OtherTok{(}\NormalTok{return e}\OtherTok{)} \OtherTok{(}\NormalTok{returnD f}\OtherTok{)}\NormalTok{ g gg }\OtherTok{=}\NormalTok{ returnD f}
\NormalTok{FreeDep⇒ }\OtherTok{(}\NormalTok{bind a k}\OtherTok{)} \OtherTok{(}\NormalTok{bindD c h}\OtherTok{)}\NormalTok{ g gg }\OtherTok{=} 
\NormalTok{    \textgreater{}\textgreater{}=Dep }\OtherTok{(}\NormalTok{g a}\OtherTok{)} \OtherTok{(}\NormalTok{gg a c}\OtherTok{)} \OtherTok{(λ}\NormalTok{ e x }\OtherTok{→}\NormalTok{ FreeDep⇒ }\OtherTok{(}\NormalTok{k e}\OtherTok{)} \OtherTok{(}\NormalTok{h e x}\OtherTok{)}\NormalTok{ g gg}\OtherTok{)}

\CommentTok{{-}{-} Kleisli composition for the free dependent monad}
\NormalTok{∘Dep }\OtherTok{:} \OtherTok{\{}\NormalTok{p }\OtherTok{:}\NormalTok{ Poly}\OtherTok{\}} \OtherTok{\{}\NormalTok{q }\OtherTok{:}\NormalTok{ Poly}\OtherTok{\}} \OtherTok{\{}\NormalTok{r }\OtherTok{:}\NormalTok{ Poly}\OtherTok{\}}
     \OtherTok{→} \OtherTok{\{}\NormalTok{s }\OtherTok{:}\NormalTok{ DepPoly p}\OtherTok{\}} \OtherTok{\{}\NormalTok{t }\OtherTok{:}\NormalTok{ DepPoly q}\OtherTok{\}} \OtherTok{\{}\NormalTok{v }\OtherTok{:}\NormalTok{ DepPoly r}\OtherTok{\}}
     \OtherTok{→} \OtherTok{\{}\NormalTok{f }\OtherTok{:}\NormalTok{ p ⇒ Free q}\OtherTok{\}} \OtherTok{\{}\NormalTok{g }\OtherTok{:}\NormalTok{ q ⇒ Free r}\OtherTok{\}}
     \OtherTok{→} \OtherTok{(}\NormalTok{⇒Dep p s }\OtherTok{(}\NormalTok{FreeDep q t}\OtherTok{)}\NormalTok{ f}\OtherTok{)}
     \OtherTok{→} \OtherTok{(}\NormalTok{⇒Dep q t }\OtherTok{(}\NormalTok{FreeDep r v}\OtherTok{)}\NormalTok{ g}\OtherTok{)}
     \OtherTok{→}\NormalTok{ ⇒Dep p s }\OtherTok{(}\NormalTok{FreeDep r v}\OtherTok{)} \OtherTok{(}\NormalTok{f ∘ g}\OtherTok{)}
\NormalTok{∘Dep ff gg a c }\OtherTok{=}\NormalTok{ FreeDep⇒ }\OtherTok{\_} \OtherTok{(}\NormalTok{ff a c}\OtherTok{)} \OtherTok{\_}\NormalTok{ gg}

\CommentTok{{-}{-} multi{-}ary Kleisli composition of }
\CommentTok{{-}{-} dependent polynomial morphisms}
\NormalTok{compDep }\OtherTok{:} \OtherTok{(}\NormalTok{U }\OtherTok{:} \DataTypeTok{Set}\OtherTok{)} \OtherTok{\{}\NormalTok{p }\OtherTok{:}\NormalTok{ Poly}\OtherTok{\}} \OtherTok{\{}\NormalTok{q }\OtherTok{:}\NormalTok{ U }\OtherTok{→}\NormalTok{ Poly}\OtherTok{\}} \OtherTok{\{}\NormalTok{r }\OtherTok{:}\NormalTok{ Poly}\OtherTok{\}}
          \OtherTok{\{}\NormalTok{s }\OtherTok{:}\NormalTok{ DepPoly p}\OtherTok{\}} \OtherTok{\{}\NormalTok{t }\OtherTok{:} \OtherTok{(}\NormalTok{u }\OtherTok{:}\NormalTok{ U}\OtherTok{)} \OtherTok{→}\NormalTok{ DepPoly }\OtherTok{(}\NormalTok{q u}\OtherTok{)\}} 
          \OtherTok{\{}\NormalTok{v }\OtherTok{:}\NormalTok{ DepPoly r}\OtherTok{\}} \OtherTok{\{}\NormalTok{f }\OtherTok{:}\NormalTok{ p ⇒ Free }\OtherTok{(}\NormalTok{sum U q}\OtherTok{)\}}
        \OtherTok{→} \OtherTok{\{}\NormalTok{g }\OtherTok{:} \OtherTok{(}\NormalTok{u }\OtherTok{:}\NormalTok{ U}\OtherTok{)} \OtherTok{→}\NormalTok{ q u ⇒ Free r}\OtherTok{\}}
        \OtherTok{→} \OtherTok{(}\NormalTok{⇒Dep p s }\OtherTok{(}\NormalTok{FreeDep }\OtherTok{(}\NormalTok{sum U q}\OtherTok{)} \OtherTok{(}\NormalTok{sumDep U q t}\OtherTok{))}\NormalTok{ f}\OtherTok{)}
        \OtherTok{→} \OtherTok{((}\NormalTok{u }\OtherTok{:}\NormalTok{ U}\OtherTok{)} \OtherTok{→}\NormalTok{ ⇒Dep }\OtherTok{(}\NormalTok{q u}\OtherTok{)} \OtherTok{(}\NormalTok{t u}\OtherTok{)} \OtherTok{(}\NormalTok{FreeDep r v}\OtherTok{)} \OtherTok{(}\NormalTok{g u}\OtherTok{))}
        \OtherTok{→}\NormalTok{ ⇒Dep p s }\OtherTok{(}\NormalTok{FreeDep r v}\OtherTok{)} \OtherTok{(}\NormalTok{comp U f g}\OtherTok{)}
\NormalTok{compDep U ff gg }\OtherTok{=}\NormalTok{ ∘Dep ff }\OtherTok{(λ} \OtherTok{(}\NormalTok{u , x}\OtherTok{)}\NormalTok{ c }\OtherTok{→}\NormalTok{ gg u x c}\OtherTok{)}

\CommentTok{{-}{-} dependent Mealy machines}
\KeywordTok{record}\NormalTok{ DepMealy }\OtherTok{(}\NormalTok{p }\OtherTok{:}\NormalTok{ Poly}\OtherTok{)} \OtherTok{(}\NormalTok{r }\OtherTok{:}\NormalTok{ DepPoly p}\OtherTok{)} \OtherTok{(}\NormalTok{m }\OtherTok{:}\NormalTok{ Mealy p}\OtherTok{)} \OtherTok{:} \DataTypeTok{Set} \KeywordTok{where}
    \KeywordTok{coinductive}
    \KeywordTok{field}
\NormalTok{        appD }\OtherTok{:} \OtherTok{(}\NormalTok{x }\OtherTok{:}\NormalTok{ p }\OtherTok{.}\NormalTok{fst}\OtherTok{)} \OtherTok{(}\NormalTok{y }\OtherTok{:}\NormalTok{ r }\OtherTok{.}\NormalTok{fst x}\OtherTok{)} 
             \OtherTok{→}\NormalTok{ r }\OtherTok{.}\NormalTok{snd x y }\OtherTok{(}\NormalTok{m }\OtherTok{.}\NormalTok{app x }\OtherTok{.}\NormalTok{fst}\OtherTok{)}
\NormalTok{               × DepMealy p r }\OtherTok{(}\NormalTok{m }\OtherTok{.}\NormalTok{app x }\OtherTok{.}\NormalTok{snd}\OtherTok{)}

\KeywordTok{open}\NormalTok{ DepMealy }\KeywordTok{public}

\CommentTok{{-}{-} specification for state invariants}
\NormalTok{Invariant }\OtherTok{:} \OtherTok{(}\NormalTok{S }\OtherTok{:} \DataTypeTok{Set}\OtherTok{)} \OtherTok{(}\NormalTok{T }\OtherTok{:}\NormalTok{ S }\OtherTok{→} \DataTypeTok{Set}\OtherTok{)} \OtherTok{→}\NormalTok{ DepPoly }\OtherTok{(}\NormalTok{State S}\OtherTok{)}
\NormalTok{Invariant S T }\OtherTok{.}\NormalTok{fst get }\OtherTok{=}\NormalTok{ ⊤}
\NormalTok{Invariant S T }\OtherTok{.}\NormalTok{fst }\OtherTok{(}\NormalTok{put s}\OtherTok{)} \OtherTok{=}\NormalTok{ T s}
\NormalTok{Invariant S T }\OtherTok{.}\NormalTok{snd get }\OtherTok{\_}\NormalTok{ s }\OtherTok{=}\NormalTok{ T s}
\NormalTok{Invariant S T }\OtherTok{.}\NormalTok{snd }\OtherTok{(}\NormalTok{put s}\OtherTok{)}\NormalTok{ t }\OtherTok{\_} \OtherTok{=}\NormalTok{ ⊤}

\CommentTok{{-}{-} verification of a state invariant}
\NormalTok{invariant }\OtherTok{:} \OtherTok{\{}\NormalTok{S }\OtherTok{:} \DataTypeTok{Set}\OtherTok{\}} \OtherTok{\{}\NormalTok{T }\OtherTok{:}\NormalTok{ S }\OtherTok{→} \DataTypeTok{Set}\OtherTok{\}}
          \OtherTok{→} \OtherTok{(}\NormalTok{s }\OtherTok{:}\NormalTok{ S}\OtherTok{)} \OtherTok{→}\NormalTok{ T s }
          \OtherTok{→}\NormalTok{ DepMealy }\OtherTok{(}\NormalTok{State S}\OtherTok{)} \OtherTok{(}\NormalTok{Invariant S T}\OtherTok{)} \OtherTok{(}\NormalTok{state{-}mealy s}\OtherTok{)}
\NormalTok{invariant s t }\OtherTok{.}\NormalTok{appD get }\OtherTok{\_} \OtherTok{=} \OtherTok{(}\NormalTok{t , }\OtherTok{(}\NormalTok{invariant s t}\OtherTok{))}
\NormalTok{invariant s t }\OtherTok{.}\NormalTok{appD }\OtherTok{(}\NormalTok{put s\textquotesingle{}}\OtherTok{)}\NormalTok{ t\textquotesingle{} }\OtherTok{=} \OtherTok{(\_}\NormalTok{ , invariant s\textquotesingle{} t\textquotesingle{}}\OtherTok{)}

\CommentTok{{-}{-} run a verified Mealy machine on a verified program}
\NormalTok{run{-}mealyD }\OtherTok{:} \OtherTok{\{}\NormalTok{p }\OtherTok{:}\NormalTok{ Poly}\OtherTok{\}} \OtherTok{\{}\NormalTok{r }\OtherTok{:}\NormalTok{ DepPoly p}\OtherTok{\}} \OtherTok{\{}\NormalTok{E }\OtherTok{:} \DataTypeTok{Set}\OtherTok{\}} \OtherTok{\{}\NormalTok{F }\OtherTok{:}\NormalTok{ E }\OtherTok{→} \DataTypeTok{Set}\OtherTok{\}} 
           \OtherTok{→} \OtherTok{\{}\NormalTok{e }\OtherTok{:}\NormalTok{ Free p E}\OtherTok{\}} \OtherTok{\{}\NormalTok{m }\OtherTok{:}\NormalTok{ Mealy p}\OtherTok{\}} 
           \OtherTok{→}\NormalTok{ FreeDep p r E F e }\OtherTok{→}\NormalTok{ DepMealy p r m }
           \OtherTok{→} \OtherTok{(}\NormalTok{F }\OtherTok{(}\NormalTok{run{-}mealy e m }\OtherTok{.}\NormalTok{fst}\OtherTok{)} 
\NormalTok{             × DepMealy p r }\OtherTok{(}\NormalTok{run{-}mealy e m }\OtherTok{.}\NormalTok{snd}\OtherTok{))}
\NormalTok{run{-}mealyD }\OtherTok{(}\NormalTok{returnD f}\OtherTok{)}\NormalTok{ mm }\OtherTok{=} \OtherTok{(}\NormalTok{f , mm}\OtherTok{)}
\NormalTok{run{-}mealyD }\OtherTok{(}\NormalTok{bindD c h}\OtherTok{)}\NormalTok{ mm }\OtherTok{=} 
    \KeywordTok{let} \OtherTok{(}\NormalTok{d , mm\textquotesingle{}}\OtherTok{)} \OtherTok{=}\NormalTok{ mm }\OtherTok{.}\NormalTok{appD }\OtherTok{\_}\NormalTok{ c }\KeywordTok{in} 
\NormalTok{    run{-}mealyD }\OtherTok{(}\NormalTok{h }\OtherTok{\_}\NormalTok{ d}\OtherTok{)}\NormalTok{ mm\textquotesingle{}}

\CommentTok{{-}{-} apply a verified program to a verified Mealy }
\CommentTok{{-}{-} machine to obtain a new verified Mealy machine}
\NormalTok{prog→mealyD }\OtherTok{:} \OtherTok{\{}\NormalTok{p q }\OtherTok{:}\NormalTok{ Poly}\OtherTok{\}} \OtherTok{\{}\NormalTok{r }\OtherTok{:}\NormalTok{ DepPoly p}\OtherTok{\}} \OtherTok{\{}\NormalTok{s }\OtherTok{:}\NormalTok{ DepPoly q}\OtherTok{\}} 
            \OtherTok{→} \OtherTok{\{}\NormalTok{f }\OtherTok{:}\NormalTok{ p ⇒ Free q}\OtherTok{\}} \OtherTok{\{}\NormalTok{m }\OtherTok{:}\NormalTok{ Mealy q}\OtherTok{\}}
            \OtherTok{→} \OtherTok{(}\NormalTok{⇒Dep p r }\OtherTok{(}\NormalTok{FreeDep q s}\OtherTok{)}\NormalTok{ f}\OtherTok{)} \OtherTok{→}\NormalTok{ DepMealy q s m}
            \OtherTok{→}\NormalTok{ DepMealy p r }\OtherTok{(}\NormalTok{prog→mealy f m}\OtherTok{)}
\NormalTok{prog→mealyD ff mm }\OtherTok{.}\NormalTok{appD a c }\OtherTok{=} 
    \KeywordTok{let} \OtherTok{(}\NormalTok{b , mm\textquotesingle{}}\OtherTok{)} \OtherTok{=}\NormalTok{ run{-}mealyD }\OtherTok{(}\NormalTok{ff a c}\OtherTok{)}\NormalTok{ mm }\KeywordTok{in} 
    \OtherTok{(}\NormalTok{b , }\OtherTok{(}\NormalTok{prog→mealyD ff mm\textquotesingle{}}\OtherTok{))}

\CommentTok{{-}{-} closure of DepMealy under binary coproducts}
\OtherTok{\_}\NormalTok{depMealy⊕}\OtherTok{\_} \OtherTok{:} \OtherTok{\{}\NormalTok{p q }\OtherTok{:}\NormalTok{ Poly}\OtherTok{\}} \OtherTok{\{}\NormalTok{r }\OtherTok{:}\NormalTok{ DepPoly p}\OtherTok{\}} \OtherTok{\{}\NormalTok{s }\OtherTok{:}\NormalTok{ DepPoly q}\OtherTok{\}}
            \OtherTok{→} \OtherTok{\{}\NormalTok{m₁ }\OtherTok{:}\NormalTok{ Mealy p}\OtherTok{\}} \OtherTok{\{}\NormalTok{m₂ }\OtherTok{:}\NormalTok{ Mealy q}\OtherTok{\}}
            \OtherTok{→}\NormalTok{ DepMealy p r m₁ }\OtherTok{→}\NormalTok{ DepMealy q s m₂ }
            \OtherTok{→}\NormalTok{ DepMealy }\OtherTok{(}\NormalTok{p ⊕ q}\OtherTok{)} \OtherTok{(}\NormalTok{depPoly⊕ r s}\OtherTok{)} \OtherTok{(}\NormalTok{m₁ mealy⊕ m₂}\OtherTok{)}
\OtherTok{\_}\NormalTok{depMealy⊕}\OtherTok{\_}\NormalTok{ m₁ m₂ }\OtherTok{.}\NormalTok{appD }\OtherTok{(}\NormalTok{inl a}\OtherTok{)}\NormalTok{ c }\OtherTok{=} 
    \KeywordTok{let} \OtherTok{(}\NormalTok{b , m\textquotesingle{}}\OtherTok{)} \OtherTok{=}\NormalTok{ m₁ }\OtherTok{.}\NormalTok{appD a c }\KeywordTok{in}
    \OtherTok{(}\NormalTok{b , }\OtherTok{(}\NormalTok{m\textquotesingle{} depMealy⊕ m₂}\OtherTok{))}
\OtherTok{\_}\NormalTok{depMealy⊕}\OtherTok{\_}\NormalTok{ m₁ m₂ }\OtherTok{.}\NormalTok{appD }\OtherTok{(}\NormalTok{inr c}\OtherTok{)}\NormalTok{ d }\OtherTok{=}
    \KeywordTok{let} \OtherTok{(}\NormalTok{e , m\textquotesingle{}}\OtherTok{)} \OtherTok{=}\NormalTok{ m₂ }\OtherTok{.}\NormalTok{appD c d }\KeywordTok{in}
    \OtherTok{(}\NormalTok{e , }\OtherTok{(}\NormalTok{m₁ depMealy⊕ m\textquotesingle{}}\OtherTok{))}
\end{Highlighting}
\end{Shaded}

\end{document}